\newif\ifconfver
\confvertrue        

\newif\ifplainver  
\plainvertrue

\newif\ifhide  
\hidetrue

\ifplainver
    \confverfalse   
\fi

\ifconfver
     \documentclass[10pt,twocolumn,twoside]{IEEEtran}
\else
    \ifplainver
        \documentclass[11pt]{article}
        \usepackage{fullpage}
    \else
        \documentclass[12pt,draftcls,onecolumn]{IEEEtran}
    \fi
\fi

\usepackage{etoolbox}%
\usepackage{xpatch}
\usepackage{blindtext}
\usepackage{tocloft}%
\newlength{\articlesectionshift}%
\let\LaTeXStandardSection\section
\let\LaTeXStandardTheSection\thesection

\let\LaTeXStandardTheSubSubSection\thesubsubsection
\let\LaTeXStandardTheParagraph\theparagraph

\makeatletter
\newcounter{titlecounter}

\xpretocmd{\maketitle}{\ifnumgreater{\value{titlecounter}}{1}}{\clearpage}{}{} 
\xpatchcmd{\maketitle}{\let\maketitle\relax\let\@maketitle\relax}{\refstepcounter{titlecounter}\begingroup
  \addtocontents{toc}{\begingroup\addtolength{\cftsecindent}{-\articlesectionshift}}%
  \addcontentsline{toc}{section}{\protect{\numberline{\thetitlecounter}{\@title~ \@author}}}%
  \addtocontents{toc}{\endgroup}
}{%
  \typeout{Patching was successful}
}{%
  \typeout{patching failed}
}%

\def\@IEEEdestroythesectionargument#1{\LaTeXStandardSection{#1}}%

\xapptocmd{\maketitle}{%
\renewcommand{\thesection}{\LaTeXStandardTheSection}%
\renewcommand{\thesubsection}{\LaTeXStandardTheSubSection}%
\renewcommand{\thesubsubsection}{\LaTeXStandardTheSubSubSection}%
\renewcommand{\theparagraph}{\LaTeXStandardTheParagraph}%
}{}{}%

\@addtoreset{section}{titlecounter}

\usepackage{calc,amsfonts,amssymb,amsmath,bm,url,color,theorem,graphicx,cite}
\usepackage{psfrag,float}
\usepackage{algorithm}
\usepackage{algorithmic}
\usepackage{soul}
\usepackage{enumerate}
\usepackage{bbm}
\usepackage{shortcuts_OPT}
\usepackage{multirow}

\usepackage{eqparbox}

\definecolor{orange}{RGB}{255,107,0}

\def\red{\color{red}}

\newtheorem{Fact}{Fact}
\newtheorem{Lemma}{Lemma}
\newtheorem{Prop}{Proposition}
\newtheorem{Theorem}{Theorem}

\newtheorem{Asm}{Assumption}
\theorembodyfont{\rmfamily}

\newtheorem{Remark}{Remark}

\newcommand\bq{\ensuremath{{\bm q}}}
\newcommand\bx{\ensuremath{{\bm x}}}
\newcommand\by{\ensuremath{{\bm y}}}
\newcommand\bG{\ensuremath{{\bm G}}}

\newcommand\bH{\ensuremath{{\bm H}}}

\newcommand\be{\ensuremath{{\bm e}}}
\newcommand\bz{\ensuremath{{\bm z}}}
\newcommand\bp{\ensuremath{{\bm p}}}

\newcommand\bR{\ensuremath{{\bm R}}}

\newcommand\bX{\ensuremath{{\bm X}}}
\newcommand\bZ{\ensuremath{{\bm Z}}}

\newcommand\ba{\ensuremath{{\bm a}}}

\newcommand\bA{\ensuremath{{\bm A}}}

\newcommand\bb{\ensuremath{{\bm b}}}
\newcommand\bg{\ensuremath{{\bm g}}}
\newcommand\bB{\ensuremath{{\bm B}}}
\newcommand\blam{\ensuremath{{\bm \lambda}}}

\newcommand\bF{\ensuremath{{\bm F}}}

\newcommand\bd{\ensuremath{{\bm d}}}
\newcommand\bD{\ensuremath{{\bm D}}}
\newcommand\bu{\ensuremath{{\bm u}}}
\newcommand\bv{\ensuremath{{\bm v}}}
\newcommand\bSig{\ensuremath{{\bm \Sigma}}}
\newcommand\bsig{\ensuremath{{\bm \sigma}}}

\newcommand\bPhi{\ensuremath{{\bm \Phi}}}
\newcommand\bPsi{\ensuremath{{\bm \Psi}}}
\newcommand\btheta{\ensuremath{{\bm \theta}}}

\newcommand\bYh{\ensuremath{{\bm Y}_{\rm H}}}
\newcommand\bYm{\ensuremath{{\bm Y}_{\rm M}}}

\newcommand\bVh{\ensuremath{{\bm V}_{\rm H}}}
\newcommand\bVm{\ensuremath{{\bm V}_{\rm M}}}
\newcommand\Lh{\ensuremath{L_{\rm H}}}
\newcommand\Mm{\ensuremath{M_{\rm M}}}

\newcommand\bY{\ensuremath{{\bm Y}}}
\newcommand\bV{\ensuremath{{\bm V}}}
\newcommand\bU{\ensuremath{{\bm U}}}
\newcommand\bs{\ensuremath{{\bm s}}}
\newcommand\bS{\ensuremath{{\bm S}}}

\newcommand{\Rbb}{\mathbb{R}}

\newcommand{\setA}{\mathcal{A}}
\newcommand{\setX}{\mathcal{X}}

\newcommand{\setU}{\mathcal{U}}

\newcommand{\setS}{\mathcal{S}}

\newcommand{\setL}{\mathcal{L}}

\newcommand{\Diag}{\mathrm{Diag}}

\newcommand\bQ{\ensuremath{{\bm Q}}}
\newcommand\bLam{\ensuremath{{\bm \Lambda}}}

\newcommand{\bzero}{{\bm 0}}
\newcommand{\bone}{{\bm 1}}
\newcommand{\bI}{{\bm I}}

\newcommand\vvec{\ensuremath{{\rm vec}}}
\newcommand\mmat{\ensuremath{{\rm mat}}}
\newcommand\bigO{\ensuremath{{\mathcal{O}}}}

\newcommand{\lammax}{\lambda_{\rm max}}
\newcommand{\lammin}{\lambda_{\rm min}}

\newcommand{\hbetamini}{{\delta_i}}
\newcommand\hbetamin[1]{{\delta_{#1}}}

\newcommand{\hbetai}{{\hat{\beta}_i}}

\newcommand{\aalp}{{\alpha_i}}
\newcommand\aaalp[1]{{\alpha_{#1}}}
\newcommand{\setI}{\mathcal{I}}

\newcommand\indfn[1]{{{\mathbbm 1}_{#1}}}
\newcommand\prox{\ensuremath{{\sf prox}}}
\newcommand\LO{\ensuremath{{\sf LO}}}

\newcommand\bxkil{\bm x^{k,i,\ell}}
\newcommand\bxki[1]{\bm x^{k,i,#1}}
\newcommand\bxk[2]{\bm x^{k,#1,#2}}
\newcommand\dF[1]{\Delta_{#1}}
\newcommand\dx[1]{e_{#1}}
\newcommand\tdx[1]{\tilde{e}_{#1}}

\usepackage{tikz}
\usetikzlibrary{arrows.meta}
\usetikzlibrary{decorations}
\usetikzlibrary{calc}
\usetikzlibrary{shapes.geometric}
\usetikzlibrary{external}

\begin{document}

\bibliographystyle{IEEEtran}

\newcommand{\papertitle}{
Hybrid Inexact BCD for Coupled Structured Matrix Factorization in  Hyperspectral Super-Resolution
}

\newcommand{\paperabstract}{
    This paper develops a first-order optimization method for coupled structured matrix factorization (CoSMF) problems that arise in the context of hyperspectral super-resolution (HSR) in remote sensing.
    To best leverage the problem structures for computational efficiency, we introduce a hybrid inexact block coordinate descent (HiBCD) scheme wherein one coordinate is updated via the fast proximal gradient (FPG) method, while another via the Frank-Wolfe (FW) method.
    The FPG-type methods are known to take less number of iterations to converge, by numerical experience, while the FW-type methods can offer lower per-iteration complexity in certain cases;
    and we wish to take the best of both.
    We show that the limit points of this HiBCD scheme are stationary.
    Our proof treats HiBCD as an optimization framework for a class of multi-block structured optimization problems, and our stationarity claim is applicable not only to CoSMF but also to many other problems.
    Previous optimization research showed the same stationarity result for inexact block coordinate descent with either FPG or FW updates only.
    Numerical results indicate that the proposed HiBCD scheme is computationally much more efficient than the state-of-the-art CoSMF schemes in HSR.
}


\ifplainver


\title{\papertitle}

    \author{
    Ruiyuan Wu$^\dag$, Hoi-To Wai$^\ddag$, and Wing-Kin Ma$^\dag$ \\ ~ \\
    $^\dag$Department of Electronic Engineering, The Chinese University of Hong Kong, \\
    Hong Kong SAR of China \\ ~ \\
    $^\ddag$Department of Systems Engineering and Engineering Management, \\ The Chinese University of Hong Kong,
    Hong Kong SAR of China
    }

    \maketitle

    \begin{abstract}
    \paperabstract
    \end{abstract}

\else
    \title{\papertitle}

    \ifconfver \else {\linespread{1.1} \rm \fi

    \author{Ruiyuan Wu, Hoi-To Wai, and Wing-Kin Ma
    }

    \maketitle

    \ifconfver \else
        \begin{center} \vspace*{-2\baselineskip}
        \end{center}
    \fi

    \begin{abstract}
    \paperabstract
    \end{abstract}


    \begin{IEEEkeywords}\vspace{-0.0cm}
        Hyperspectral super-resolution, coupled matrix factorization, non-convex optimization
    \end{IEEEkeywords}

    \ifconfver \else \IEEEpeerreviewmaketitle} \fi

 \fi

\ifconfver \else
    \ifplainver \else
        \newpage
\fi \fi

\section{Introduction}

{\let\thefootnote\relax\footnotetext{This research was supported by project \#MMT-8115059 of the Shun Hing
    Institute of Advanced Engineering, The Chinese University of Hong Kong. The conference version of this paper appeared in ICASSP 2018.}}
Consider the following problem:
We have a data matrix $\bX$ to sense, and the observation is a pair of column and row decimated versions of $\bX$
\begin{equation} \label{eq:Ym_Yh}
\bYm = \bF \bX, \quad \bYh = \bX\bG,
\end{equation}
where $\bF$ and $\bG$ are fat and tall, respectively (resp.).
Here, the number of observations (or the sum of the numbers of elements of $\bYm$ and $\bYh$) is less than that of the unknowns (or the number of elements of $\bX$).
If one seeks to find $\bX$ by writing \eqref{eq:Ym_Yh} as a linear system, and then solving it, the solution will be non-unique---and the true $\bX$ will not be distinguished.
On the other hand, the problem allows us to assume that $\bX$ takes a low-rank structure $\bX = \bA \bS$ for some tall $\bA$ and fat $\bS$.
Also, $\bA$ and $\bS$ exhibit certain structures, e.g., non-negativeness.
It therefore makes sense to consider a coupled structured matrix factorization (CoSMF)
\begin{equation} \label{eq:CoSMF_zero}
\min_{\bA \in \setA, \bS \in \setS} ~
\tfrac{1}{2} \| \bYm - \bF \bA \bS \|_F^2 + \tfrac{1}{2} \| \bYh - \bA \bS \bG \|_F^2,
\end{equation}
for some structure-specifying set $\setA \times \setS$, to recover $\bX$.

\begin{figure*}[htb]
    \centering
    \includegraphics[width=0.9\linewidth]{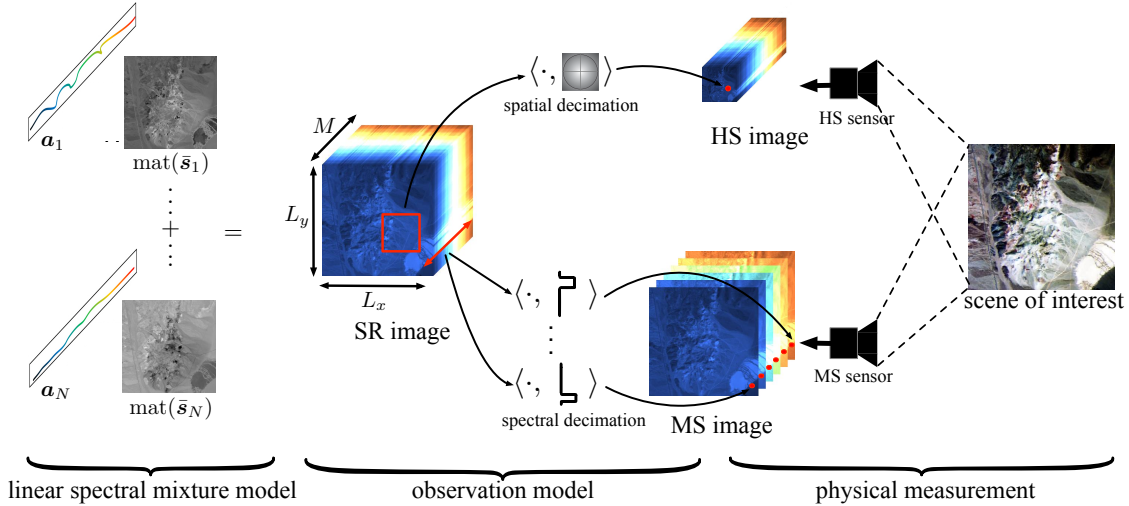}
    \caption{Illustration of the data model.}
    \label{fig:datamodel}
\end{figure*}

The above problem is a simplified description of a strongly motivated problem in hyperspectral imaging for remote sensing---namely, {\em hyperspectral super-resolution} (HSR) \cite{loncan2015hyperspectral,yokoya2017hyperspectral}.
Hyperspectral (HS) images have rich spectral contents, and such property has been extensively utilized in a myriad of remote sensing applications.
HSR seeks to enhance the spatial resolution of the HS image with the aid of another image, namely, a multispectral (MS) image, which has higher spatial resolution than the HS image, but has low spectral resolution.
In HSR, $\bYm$ and $\bYh$ in \eqref{eq:Ym_Yh} represent an MS image and an HS image, resp., in a spectral-spatial matrix form.
The problem is to construct a super-resolution (SR) image $\bX$---whose spectral and spatial resolutions are identical to those of the HS and MS images, resp.---from the MS-HS image pair.
As illustrated in Fig.~\ref{fig:datamodel}, and as will be elucidated later, the MS and HS images can be modeled as spectral and spatial decimations of the SR images, resp.
This gives rise to the model \eqref{eq:Ym_Yh}, and subsequently, the CoSMF formulation \eqref{eq:CoSMF_zero}.
Since acquiring images of both high spectral and spatial resolutions is 
difficult, if not impossible, from the optical sensing perspective, this MS-aided HSR solution is very intriguing.
In remote sensing we have recently observed rapidly growing activities on HSR,
{particularly, those under the CoSMF paradigm.
In this paper we will narrow down our scope to CoSMF, and before we do so we should note that HSR can also be tackled by image enhancement techniques  such as pan-sharpening \cite{loncan2015hyperspectral,yokoya2017hyperspectral} and deep learning \cite{palsson2017multispectral} (also \cite{lanaras2018super} for a  related work).}


Herein we concisely review the CoSMF developments using a signal processing lens;
the reader are referred to
\cite{loncan2015hyperspectral,yokoya2017hyperspectral} for a comprehensive coverage.
The idea of formulating HSR as a CoSMF problem was independently introduced by Kawakami {\em et al.}~\cite{kawakami2011high} and Yokoya {\em et al.}~\cite{yokoya2012coupled}.
Particularly, Yokoya {\em et al.} considered non-negative factorization; i.e., $\setA$ and $\setS$ in \eqref{eq:CoSMF_zero} are non-negative matrix sets.
Subsequent research explored many different ways to exploit the problem structures, e.g.,
$\setA$ and $\setS$ arising from some data model \cite{lanaras2015hyperspectral,wei2016multiband}, and
the 2D spatial structure exploited through total-variation regularizaion \cite{simoes2015convex,lanaras2017hyperspectral}.
Lately it is shown that, theoretically, CoSMF can provide certain recovery guarantees \cite{liu2019hsr_recovery}.
Some research also studied decoupled CoSMF methods  \cite{simoes2015convex,hsr_recovery_ssp2018} and  dictionary learning {\cite{akhtar2014sparse,wei2015hyperspectral}}.
More recent work extends CoSMF to coupled tensor factorization  \cite{kanatsoulis2018hyperspectral,li2018fusing,prevost2019coupled}.

The CoSMF problem is, by nature, a non-convex large-scale optimization problem.
We are interested in developing an efficient optimization method that can best leverage the  problem structures of CoSMF in HSR.
State-of-the-art methods in the remote sensing literature often consider some intuitive alternating update methods \cite{yokoya2012coupled,lanaras2015hyperspectral},
or block coordinate descent (BCD) and the practical variants thereof \cite{wei2016multiband}.
We propose a hybrid inexact BCD (HiBCD) scheme,
wherein each block coordinate can be updated by either the fast proximal gradient (FPG) method or the Frank-Wolfe (FW) method.
The reason for considering such hybrid updates is that some block coordinate is better suited to one particular update method,
and this makes us wonder if we can take the best from both.
The proposed scheme is inexact in the sense that each block coordinate update needs not solve a minimization problem exactly---the requirement in standard BCD.
Relaxing exact updates gives us the flexibility to build algorithms with better computational efficiency, as exact BCD can be computationally heavy in the per-iteration sense.

This work has two key contributions.
The first is practical.
We custom-develop the HiBCD scheme for the CoSMF problem, specifically, at the implementation or computational level.
It will be shown by numerical experiments that the proposed HiBCD scheme runs many times faster than the state-of-the-art methods.
The second is theoretical.
While our HiBCD development is motivated by the HSR application,
it can conceptually be cast as an optimization framework for a class of multi-block structured optimization problems.
From such a perspective, we analyze the sufficient conditions under which HiBCD guarantees some form of convergence to a stationary point.
In the mathematical optimization literature, we have seen a rich collection of optimization frameworks that deal with the same or similar problem class covered by this work \cite{wright2015coordinate,razaviyayn2013unified,hong2017iteration,xu2013block,xu2017globally,beck2015cyclic,beck2018primal}.
None of them, however, studies hybrid schemes like we do.
Our HiBCD analysis unifies those of the alternating proximal gradient method \cite{xu2013block,xu2017globally} and the cyclic block conditional gradient (CBCG) method \cite{beck2015cyclic}, which are pure FPG and FW instances of our HiBCD scheme, resp.

The organization of this paper is as follows.
Sections~\ref{sec:prelim} and \ref{sec:prob_review} review some preliminary  concepts and the problem background, resp.
Section~\ref{sec:cosmf} describes the HiBCD scheme for CoSMF.
Section~\ref{sec:analysis_hibcd} analyzes the convergence of HiBCD.
Numerical results are provided in Section~\ref{sec:num_exp}, and conclusion is drawn in Section~\ref{sec:conclude}.

{The reader can find the source code of our HiBCD scheme at \url{https://github.com/REIYANG/HiBCD}.}

\section{Preliminaries}
\label{sec:prelim}

\subsection{Notations}

Unless specified otherwise, we will adopt the following notations.
The $i$th column of a matrix $\bX$ is denoted by $\bx_i$, while the $j$th row by $\bar{\bx}_j$;
$\bzero$ denotes an all-zero vector or matrix;
$\bone$ denotes an all-one vector or matrix;
$\be_i$ denotes a unit vector with $1$ at the $i$th entry;
$\bI_n$ denotes the $n \times n$ identity matrix;
$\bI$ denotes an identity matrix of appropriate size;
$\Diag(\bx)$ denotes a diagonal matrix with the main diagonal entries given by $x_1,\ldots,x_n$;
$\| \cdot \|$, when applied on vectors, means the Euclidean norm;
$\| \cdot \|_F, \| \cdot \|_2$ and $\| \cdot \|_*$ denote the Frobenius, spectral and nuclear norms, resp.;
$\langle \cdot, \cdot \rangle$ is the inner product;
$[ \cdot ]_{\ba}^\bb$ means that if $\by = [ \bx ]_\ba^\bb$, then $y_i = \max\{ a_i, \min\{ x_i, b_i \}\}$ for all $i$;
$\bx \geq \by$ or $\bX \geq \bY$ denotes the element-wise inequality;
$\bX \succeq \bY$ means that $\bX- \bY$ is positive semidefinite (PSD);
$\lammin(\bX)$ and $\lammax(\bX)$ denote the smallest and largest eigenvalues of $\bX$, resp.;
$\otimes$ denotes the Kronecker product;
{$\vvec( \cdot )$ and $\mmat(\cdot)$ are the vectorization and matricization operators, resp.,
i.e., $\vvec(\bX) = [~ \bx_1^\top, \cdots, \bx_m^\top ~]^\top$ and $\mmat(\vvec(\bX))= \bX$;}
we have $a \wedge b = \min\{a, b\}$.

 Also, given a function $f: \Rbb^n \rightarrow (-\infty, \infty]$, the notation
${\rm dom}(f):= \{ \bx \in \Rbb^n \mid f(\bx) < \infty \}$ denotes the domain of $f$;
the gradient of $f$ is denoted by $\nabla f$;
the gradient of a multi-block function $f(\bx_1,\ldots,\bx_m)$ with respect to (w.r.t.) the $i$th block $\bx_i$ is denoted by $\nabla_{\bx_i} f$;
the subdifferential of $f$ at $\bx$ is denoted by $\partial f(\bx)$;
$\indfn{\setX}$ denotes the indicator function of $\setX$, i.e.,
$\indfn{\setX}(\bx) = 0$ if $\bx \in \setX$, and $\indfn{\setX}(\bx) = \infty$ if $\bx \notin \setX$;
${\rm aff} \, \setX$ and ${\rm conv} \, \setX$ denote the affine and convex hulls of  $\setX$, resp.

\subsection{Lipschitz Continuity}
\label{sec:Lip_cont}

The notion of Lipschitz continuity is widely used in first-order optimization \cite{beck2017first}.
Let $f: \Rbb^n \rightarrow \Rbb$ and $\setX \subseteq \Rbb^n$.
We say that $f$ is $\beta$-Lipschitz continuous on $\setX$ if
\begin{equation} \label{eq:lip_def}
| f(\bx) - f(\by) | \leq \beta \| \bx - \by \|, \quad \text{for all $\bx, \by \in \setX$.}
\end{equation}
The accompanying parameter $\beta$ is called a Lipschitz constant of $f$ on $\setX$.
Additionally, $\beta$ is said to be tight if it is the smallest $\beta$ such that \eqref{eq:lip_def} holds.
When there is no importance with specifying $\beta$, we may simply say that $f$ is Lipschitz continuous on $\setX$.
If $f$ is convex and proper, then it is Lipschitz continuous on any compact  $\setX$ \cite[Proposition 2.4.2]{lange2016mm}.

Similarly, a differentiable function $f$ is said to have $\beta$-Lipschitz, or simply Lipschitz, continuous gradient on $\setX$ if
\begin{equation} \label{eq:lip_grad_def}
\| \nabla f(\bx) - \nabla f(\by) \| \leq \beta \| \bx - \by \|, \quad \text{for all $\bx, \by \in \setX$.}
\end{equation}
Here $\beta$ is called a Lipschitz constant of $\nabla f$ on $\setX$, and it is said to be tight if it is the smallest $\beta$ such that \eqref{eq:lip_grad_def} holds.
If $f$ is twice differentiable, then it has Lipschitz continuous gradient on any compact $\setX$.

A related concept is weak convexity \cite{drusvyatskiy2018proximal}. We say that $f$ is $\rho$-weakly convex on a convex $\setX$ if $f( \bx) + \frac{\rho}{2} \| \bx \|^2$ is convex on $\setX$. If $f$ is convex, then it is $0$-weakly convex; if $f$ has $\beta$-Lipschitz continuous gradient, then it is $\beta$-weakly convex.

Another related concept is quadratic upper bound approximation, or sufficient descent, for differentiable $f$, given by
\begin{equation} \label{eq:qub}
f(\by) \leq f(\bx) + \langle \nabla f(\bx), \by - \bx \rangle + \frac{\beta}{2} \| \by - \bx \|^2,
\end{equation}
for some $\beta$.
If $f$ has $\beta'$-Lipschitz continuous gradient on $\setX$,
then, for $\beta \geq \beta'$, \eqref{eq:qub} holds for any $\bx, \by \in \setX$.
It is subtle to note that, in some cases, \eqref{eq:qub} can hold even if $\beta$ is smaller than the tight Lipschitz constant.
One such case is shown below.
\begin{Fact} \label{fac:quad_qub}
    Let $f(\bx) = p + \bq^\top \bx + \bx^\top \bR \bx/2$, where $\bR$ is symmetric PSD.
    Let $\setX$ be a set of non-empty relative interior, and represent its affine hull by ${\rm aff} \, \setX= \{ \bx = \bPhi \bm \xi + \bd  \mid \bm \xi \in \Rbb^r \}$ for some semi-orthogonal $\bPhi \in \Rbb^{n \times r}$ and  $\bd \in \Rbb^n$.
    Then,
    \begin{enumerate}[1.]
        \item the smallest $\beta$ for which \eqref{eq:qub} holds for any $\bx, \by \in \setX$ is $\beta = \lammax(\bPhi^\top \bR \bPhi)$;
        \item the tight Lipschitz constant of $\nabla f$ on $\setX$ is $\| \bR \bPhi \|_2$, and it is true that $\| \bR \bPhi \|_2 \geq \lammax(\bPhi^\top \bR \bPhi)$.
    \end{enumerate}
\end{Fact}
{The proof of Fact~\ref{fac:quad_qub} is relegated to
Appendix \ref{app:fac:quad_qub}.}

The above notions apply to multi-block functions.
Let $\bx=(\bx_1,\ldots,\bx_m)$.
Let $\setX = \setX_1 \times \cdots \times \setX_m$ where $\setX_i$ corresponds to $\bx_i$.
Denote $\bx_{-i}= ( \bx_j )_{j=1, j\neq i}^m$,
and write the set of $\bx_{-i}$ as $\setX_{-i}$.
We say that $f$ has block-wise Lipschitz continuous gradient on $\setX$ with parameters $\beta_1,\ldots,\beta_m$ if, for each $i$,
\[
\| \nabla_{\bx_i}f(\bx) - \nabla_{\bx_i} f(\by) \| \leq \beta_i \| \bx_i - \by_i \|,
\]
for all $\bx, \by \in \setX$ with $\bx_{-i}= \by_{-i}$.
The parameter $\beta_i$ is called a block-wise Lipschitz constant of $\nabla_{\bx_i}f$ on $\setX$.
Assuming convex $\setX_i$'s, $f$ is said to be block-wise weakly convex on $\setX$ with parameters $\rho_1,\ldots,\rho_m$ if, for each $i$, $f(\bx) + \frac{\rho_i}{2} \| \bx_i \|^2$ is convex in $\bx_i \in \setX_i$ for any $\bx_{-i} \in \setX_{-i}$.
Under the block-wise Lipschitz continuous gradient condition, it holds that
\[
f(\by) \leq f(\bx) + \langle \nabla_{\bx_i} f(\bx), \by_i - \bx_i \rangle + \frac{\beta_i}{2} \| \by_i - \bx_i \|^2
\]
for all $\bx, \by \in \setX$ with $\bx_{-i}= \by_{-i}$; and that
 $f$ is block-wise weakly convex on $\setX$ with parameters $\beta_1,\ldots,\beta_m$.
If $f$ has Lipschitz continuous gradient on $\setX$, it has block-wise Lipschitz continuous gradient on $\setX$.

\subsection{Proximal Operators and Linear Optimization Oracles}
\label{sec:prox_n_LO}

The proximal operator and linear optimization (LO) oracle are basic building blocks in first-order optimization.
Let $h: \Rbb^n \rightarrow \Rbb$ be convex, closed and proper.
The proximal operator and LO oracle of $h$ are defined, resp., as
\begin{align}
\prox_h(\bx) & = \arg \min_{\by \in \Rbb^n} \, \tfrac{1}{2} \| \by - \bx \|^2 + h(\by),
\label{eq:prox_def}
\\
\LO_h(\bg) & \in \arg \min_{\by \in \Rbb^n} \, \langle \bg, \by \rangle + h(\by).
\label{eq:LO_def}
\end{align}
For notational convenience, we also define
\begin{align*}
\prox_\setX(\bx) & = \prox_{\indfn{\setX}}(\bx)= \arg \min_{\by \in \setX} \, \tfrac{1}{2} \| \by - \bx \|^2, \\
\LO_\setX(\bg) & = \LO_{\indfn{\setX}}(\bg) \in \arg \min_{\by \in \setX} \, \langle \bg, \by \rangle,
\end{align*}
where $\setX$ is closed and convex, and additionally, for the LO case, compact.
Note that $\prox_\setX$ is identical to the projection onto $\setX$.
We are interested in cases
for which
\eqref{eq:prox_def} and \eqref{eq:LO_def} are
efficiently computable.
There is a rich list of such $h$'s \cite{beck2017first,pmlr-v28-jaggi13},
and here we name some that will be used in this work.
\begin{enumerate}[1.]
    \item Let $\setX = [ 0, 1 ]^n$. We have $\prox_\setX(\bx)= [ \bx ]_\bzero^\bone$ and
     $\LO_\setX(\bg) = \bz$, where
     $z_i = 1$ if $g_i < 0$, and $z_i= 0$ if $g_i \geq 0$.

    \item Let $\setX$ be the unit simplex on $\Rbb^n$, i.e.,
    \begin{equation} \label{eq:Un}
    \setX = \setU^n := \{ \bx \in \Rbb^n \mid \bx \geq \bzero, \bone^\top \bx = 1 \}.
    \end{equation}
    We do not have an explicit expression for $\prox_\setX$, but there exist algorithms that  compute the solution to the problem in \eqref{eq:prox_def} in $\bigO(n \log(n))$ operations \cite{duchi2008efficient}.
    Also we have $\LO_\setX(\bg) = \be_j$, where $j$ is such that $g_j = \min_{i=1,\ldots,n} g_i$.

    \item Let $\setX$ be the nuclear norm ball
    \[
    \setX= \{ \bX \in \Rbb^{m \times n} \mid \| \bX \|_* \leq \gamma \},
    \]
    where $\gamma > 0$.
    Assume $m \leq n$ without loss of generality.
    Denote the singular value decomposition (SVD) of a given matrix $\bX \in \Rbb^{m \times n}$ by $\bX = \bU [~ \bSig ~ \bzero ~] \bV^\top$, where $\bSig = \Diag(\bsig)$, and $\bsig= [~ \sigma_1, \ldots, \sigma_m ~]^\top$ contains the singular values.
    We have $\prox_\setX(\bX) = \bU [~ \tilde{\bSig} ~ \bzero ~] \bV^\top$, where $\tilde{\bSig}= \Diag(\tilde{\bsig})$, and $\tilde{\bsig} = \gamma\cdot  \prox_{\setU^m}( \bsig/\gamma)$ \cite{agarwal2012fast}.
    This proximal operation requires full SVD and unit-simplex projection, which take a complexity of $\bigO(m^2 n + m \log(m))$ in total.
    The LO oracle is $\LO_\setX(\bX) = \gamma \bu_1 \bv^\top_1$, where $\bu_1$ and $\bv_1$ are the principal left and right singular vectors of $\bX$, resp.
    For large-scale $\bX$, some methods, such as the Lanczos method and the power method, can compute $\bu_1$ and $\bv_1$ much more efficiently than the full SVD.
\end{enumerate}


\section{Problem Statement}
\label{sec:prob_review}

\subsection{Model}

Fig.~\ref{fig:datamodel} illustrates the problem scenario.
The SR image we seek to construct is a tensor whose $(i,j,k)$th element $x_{ijk}$ is the reflectance of the scene at spectral band indexed by $i$ and at spatial position indexed by $(j,k)$.
The number of spectral bands of this SR image is denoted by $M$ (typically about $100$ to $200$), and the image size by $L_y \times L_x$.
We represent the SR image in a spectral-spatial form by defining a matrix $\bX \in \Rbb^{M \times L}$ whose elements are given by $x_{i,j+(k-1)L_y} = x_{ijk}$ for all $i,j,k$.
Here we denote $L= L_x L_y$,
and each column $\bx_i$ of $\bX$ describes the spectral response of a pixel.
The SR image is observed by an MS sensor and an HS sensor.
The spectral-spatial matrix of the MS image is modeled as
\begin{equation} \label{eq:Ym_model}
\bYm = \bF \bX + \bVm,
\end{equation}
where $\bYm \in \Rbb^{\Mm \times L}$ is the spectral-spatial matrix of the MS image;
$\Mm < M$ is the number of MS bands (typically about $4$ to $10$);
$\bF \in \Rbb^{\Mm \times M}$ is a spectral decimation response;
$\bVm$ is noise.
Eq.~\eqref{eq:Ym_model} implements the process of reducing a large number of fine spectral contents into several coarse-band spectral contents by means of band averaging.
The spectral pixels of the HS image, on the other hand, are modeled as
\begin{equation} \label{eq:byhi}
\by_{{\rm H},i} = \sum_{j \in \setL_i}\bx_j g_{ji} + \bv_{{\rm H},i}
\quad i=1,\ldots,\Lh,
\end{equation}
where each $\by_{{\rm H},i}$ is an HS pixel;
$\Lh < L$ is the number of HS pixels;
$\setL_i \subset \{ 1,\ldots, L \}$ indicates a neighborhood of SR spectral pixels that form $\by_{{\rm H},i}$;
$\bv_{{\rm H},i}$ is noise;
$\{ g_{ji} \}_{j \in \setL_i}$ is a spatial decimation response.
Eq. \eqref{eq:byhi} implements the process of spatial content reduction, through spatial blurring and downsampling.
For convenience we rewrite \eqref{eq:byhi} as
\begin{equation} \label{eq:Yh_model}
\bYh = \bX \bG + \bVh,
\end{equation}
where $\bYh = [~  \by_{{\rm H},1}, \ldots, \by_{{\rm H},\Lh} ~]$;
$\bVh = [~  \bv_{{\rm H},1}, \ldots, \bv_{{\rm H},\Lh} ~]$;
$\bG \in \Rbb^{L \times \Lh}$ is such that
\begin{equation} \label{eq:XG}
\bX \bG  = [~ \bX_{\setL_1} \bg_1, \ldots,  \bX_{\setL_{\Lh}} \bg_{\Lh}  ~ ],
\end{equation}
with $\bg_i \in \Rbb^{|\setL_i|}$ being the concatenation of $\{ g_{ji} \}_{j \in \setL_i}$.
Note that the sets $\setL_i$'s generally have overlaps.
In our work, the spatial decimation response $\bg_i$ usually corresponds to a finite-width blurring kernel (e.g., a $11 \times 11$ truncated 2D Gaussian kernel).
Thus, we assume that every $|\setL_i|$ is small.

The SR image is posited to follow the popularly-used linear spectral mixture model \cite{Jose12}
\begin{equation} \label{eq:lmm}
\bX = \bA \bS,
\end{equation}
for some $\bA \in \Rbb^{M \times N}, \bS \in \Rbb^{N \times L}$ and for some positive integer $N < \min\{ M, L\}$.
Here, each column $\ba_i$ of $\bA$ describes the spectral response of a distinct material, or endmember;
each column $\bs_i$ of $\bS$ describes the proportions, or abundances, of the various materials that appear in pixel $i$;
$N$ is the number of materials.
It is assumed that every $\bs_i$ lies in the unit simplex $\setU^N$ in \eqref{eq:Un},
and that $\bzero \leq \bA \leq \bone$.
In addition, each row $\bar{s}_j$ of $\bS$ describes the abundance map of a material.
We may assume that the abundance maps possess low rank or spatial smoothness characteristics, by the same spirit in low-rank and total-variation image denoising methods.

\subsection{Coupled Structured Matrix Factorization}

Under the preceding data model, an approach to construct the SR image $\bX$ from the MS-HS image pair $(\bYm,\bYh)$ is to consider a CoSMF
\begin{equation} \label{eq:CoSMF}
\min_{\bA \in \setA, \bS \in \setS} ~ f(\bA,\bS):= f_{\rm M}(\bA,\bS) + f_{\rm H}(\bA,\bS),
\end{equation}
where $\setA$ and $\setS$ are the constraint sets of $\bA$ and $\bS$, resp.;
\[
f_{\rm M}(\bA,\bS)= \tfrac{1}{2} \| \bYm - \bF \bA \bS \|_F^2,  ~
f_{\rm H}(\bA,\bS)= \tfrac{1}{2} \| \bYh - \bA \bS \bG \|_F^2.
\]
Coupled non-negative matrix factorizaton (CNMF), a pioneering CoSMF method for HSR, considers both $\setA$ and $\setS$ as non-negative matrix sets \cite{yokoya2012coupled}.
Here we are interested in
\begin{equation} \label{eq:setAS}
\setA = [0,1]^{M \times N}, ~ \setS =
\{ \bS\in \Rbb^{N \times L} \mid \bs_i \in \setU^N, ~ \forall i \},
\end{equation}
which use the structures of the linear spectral mixture model.
For convenience, the formulation \eqref{eq:CoSMF}--\eqref{eq:setAS} will be called the {\em plain CoSMF}.
Plain CoSMF was introduced in \cite{wei2016multiband,lanaras2015hyperspectral},
and it is recently shown to possess certain recovery guarantees \cite{liu2019hsr_recovery}.
We will be interested in developing an efficient first-order scheme for plain CoSMF.

There are many variants with the plain CoSMF.
Among them, one idea is to exploit the 2D spatial structures of the abundance maps \cite{simoes2015convex,lanaras2017hyperspectral}.
To demonstrate the potential of the optimization scheme to be proposed,
we will consider one such spatial structure-exploiting formulation where
\begin{equation} \label{eq:setS_NNB}
\setA = [0,1]^{M \times N}, ~
\setS =  \{ \bS\in \Rbb^{N \times L} \mid \| \mmat(\bar{\bs}_i) \|_* \leq \tau_i, ~
\forall i \},
\end{equation}
for some pre-fixed $\tau_i > 0$.
{Here, $\mmat$ folds $\bar{\bs}_i$ into an $L_x \times L_y$ 2D spatial matrix, which is an abundance map as illustrated in Fig.~\ref{fig:datamodel}.}
In this formulation, the nuclear-norm ball is used to promote low-rank structures of the abundance maps.
We will call
the formulation
\eqref{eq:CoSMF} and \eqref{eq:setS_NNB} the {\em nuclear norm constrained (NNC)-CoSMF}.

\subsection{State of the Arts}

The CoSMF problem \eqref{eq:CoSMF}
is non-convex and of large scale.
One approach of attacking it is to
apply
the following alternating scheme
\begin{align} \label{eq:fake_BCD}
\bA^{k+1} & \in \arg \min_{\bA \in \setA} f_{\rm H}(\bA,\bS^k), ~
\bS^{k+1} \in \arg \min_{\bS \in \setS} f_{\rm M}(\bA^{k+1},\bS)
\end{align}
where $\bA^k$ and $\bS^k$ are the iterates generated.
The CNMF algorithm in \cite{yokoya2012coupled} uses the Lee-Seung multiplicative updates to implement the minimizations in \eqref{eq:fake_BCD},
while the SupResPALM algorithm in \cite{lanaras2015hyperspectral} uses the proximal gradient updates.
To make the algorithms efficient, CNMF and SupResPALM apply limited numbers of updates in each alternating cycle---which means that they are inexact versions of the alternating scheme \eqref{eq:fake_BCD}.
Empirically, alternating algorithms such as CNMF and SupResPALM were reported to yield good recovery performance.
Theoretically, it is presently not known whether the alternating scheme \eqref{eq:fake_BCD} and the variants thereof
would guarantee some form of convergence.

Another approach is to apply the alternating minimization
\begin{align} \label{eq:exact_BCD}
\bA^{k+1} & \in \arg \min_{\bA \in \setA} f(\bA,\bS^k), ~
\bS^{k+1} \in \arg \min_{\bS \in \setS} f(\bA^{k+1},\bS),
\end{align}
which is block coordinate descent (BCD) in the optimization literature.
It is known, by the two-block BCD convergence result in \cite[Corollary~2]{grippo2000convergence}, that any limit point of the iterate $(\bA^k,\bS^k)$ in \eqref{eq:exact_BCD} is a stationary point of the CoSMF problem~\eqref{eq:CoSMF}.
The FUMI algorithm in \cite{wei2016multiband} implements BCD by applying custom-made solvers to the minimizations in \eqref{eq:exact_BCD};
specifically, the authors of \cite{wei2016multiband} assume 2D circulant structures with the spatial decimation process in \eqref{eq:byhi}, and they built ADMM algorithms that exploit the aforementioned structures to efficiently compute the solutions to the problems in \eqref{eq:exact_BCD}.
The potential downside with BCD is that each exact minimization update in \eqref{eq:exact_BCD} is, inevitably, computationally heavy.
In fact, for efficient implementations, the authors of FUMI limit the number of ADMM iterations to $30$ in their experiments (see \cite[Section~V.B.1]{wei2016multiband}), which, strictly speaking, implements an inexact BCD scheme.


\section{HiBCD for CoSMF}\label{sec:cosmf}

Herein we propose a hybrid inexact BCD (HiBCD) scheme for CoSMF.
The scheme is described as follows:
\begin{subequations} \label{eq:HiBCD_CoSMF}
\begin{align} \label{eq:HiBCD_CoSMF_A}
\bA^{k+1} & = \left\{ \begin{array}{ll}
    \prox_\setA \left( \bA^k_{\sf ex} - \tfrac{1}{\hat{\beta}_A^k} \nabla_\bA f(\bA^k_{\sf ex},\bS^k)  \right),
    & {\cal FPG} \\
    \bA^k + \gamma_A^k ( \LO_\setA(\nabla_\bA f(\bA^k ,\bS^k)) - \bA^k), & {\cal FW}
    \end{array}  \right.   \\
    \label{eq:HiBCD_CoSMF_S}
\bS^{k+1} & = \left\{ \begin{array}{ll}
\prox_\setS \left( \bS^k_{\sf ex} - \tfrac{1}{\hat{\beta}_S^k} \nabla_\bS f(\bA^{k+1},\bS^k_{\sf ex} )  \right),
& {\cal FPG} \\
\bS^k + \gamma_S^k ( \LO_\setS(\nabla_\bS f(\bA^{k+1},\bS^k )) - \bS^k), & {\cal FW}
\end{array}  \right.
\end{align}
\end{subequations}
for $k=0,1,2,\ldots$
Here, ${\cal FPG}$ and ${\cal FW}$ stand for the fast proximal gradient (FPG) update and the Frank-Wolfe (FW) update, resp;
$1/\hat{\beta}_A^k$ and $1/\hat{\beta}_S^k$ are the step sizes for the FPG updates;
$\bA^k_{\sf ex}$ and $\bS^k_{\sf ex}$ are the extrapolated points, given by
\[
\bA^k_{\sf ex}= \bA^k + \alpha_k ( \bA^k -  \bA^{k-1}  ), \quad
\bS^k_{\sf ex}= \bS^k + \alpha_k ( \bS^k -  \bS^{k-1}  ),
\]
where $\{ \alpha_k \}_{k \geq 0}$, with $\alpha_k \in [0,1]$, is a pre-fixed extrapolation sequence, and $\bA^{-1} = \bA^0$, $\bS^{-1} = \bS^0$;
$\gamma_A^k, \gamma_S^k \in [0,1]$ are the step sizes for the FW updates.
A typical choice  of $\{ \alpha_k \}_{k \geq 0}$ is the FISTA extrapolation sequence 
\cite{beck2017first}:
\begin{align} \label{eq:fista_seq}
\alpha_k = \tfrac{\mu_k-1}{\mu_{k+1}}, \quad
\mu_{k+1} = \tfrac{1}{2} \left( 1 + \sqrt{1+4\mu_k} \right),
\quad \mu_0 = 1.
\end{align}
The scheme \eqref{eq:HiBCD_CoSMF} is an inexact BCD in which each exact BCD minimization in \eqref{eq:exact_BCD} is replaced by either a one-step FPG update or a one-step FW update.
The update of each block can be different, e.g., FPG for $\bA$, and FW for $\bS$.

We should explain our intuition on considering this hybrid scheme.
By numerical experience, the FPG method and its variants have been observed to yield fast convergence in terms of the number of iterations used.
This observation is not just for convex problems
\cite{beck2017first},
but also for non-convex problems \cite{xu2013block,fu2016robust,shao2019framework}.
In comparison, the FW-type methods are usually slower in convergence as revealed by empirical study.
On the other hand, there are cases in which the FW-type methods have much lower per-iteration computational costs than the FPG-type method;
such cases give rise to the opportunity for the FW-type methods to serve as a more efficient solution strategy.
In our problem, the aforementioned situation happens with the update of $\bS$.
This motivates us to consider a hybrid scheme in which the FW update for $\bS$ is used to reduce the per-iteration costs, while the FPG update for $\bA$ is applied to leverage on the fast convergence of FPG (intuitively).

In the following subsections we will describe the implementations of the FPG and FW updates in \eqref{eq:HiBCD_CoSMF}.
The convergence of the HiBCD scheme will be examined in the next section.

\subsection{The FPG Updates}\label{sec:fpg}

We first consider the FPG update of $\bS$ in \eqref{eq:HiBCD_CoSMF_S}.
We need to deal with i) the computation of the gradient $\nabla_\bS f$, ii) the operations of $\prox_\setS$, and iii) the step-size selection.
For the gradient, it can be shown that
\begin{equation} \label{eq:grad_S}
\nabla_\bS f(\bA,\bS)  = (\bF \bA)^\top ( \bF \bA \bS - \bYm ) + \bA^\top ( \bA \bS \bG - \bYh ) \bG^\top.
\end{equation}
By arranging the matrix multiplications carefully,
the above gradient can be computed in $\bigO(N (LM + \sum_{i=1}^{\Lh} |\setL_i| ))$ operations;
the spatial decimation identity in \eqref{eq:XG} is necessary for this efficient computation\footnote{An essential subroutine of computing \eqref{eq:grad_S} is with the computation of $\bS\bG$.
Using \eqref{eq:XG}, in which the structured sparsity pattern of $\bG$ is considered, we can compute $\bS\bG$ in $\bigO(N \sum_{i=1}^{\Lh} |\setL_i| ))$ operations.
This computational cost is much less than that of treating $\bS\bG$ as a generic matrix multiplication, which takes $\bigO(L N \Lh)$ operations.
Likewise, we apply the same trick to compute $[\bA^\top ( \bA \bS \bG - \bYh )] \bG^\top$ efficiently.
}.
The $\prox_\setS$ operations are adaptations of the proximal operations reviewed in Section~\ref{sec:prox_n_LO}.
If $\setS$ is the column-wise unit-simplex in \eqref{eq:setAS}, then $\prox_\setS$ is column-wise unit-simplex projection; the complexity is $\bigO(L N \log(N))$.
If $\setS$ is the row-wise nuclear-norm ball  in \eqref{eq:setS_NNB}, then $\prox_\setS$ is row-wise projection onto the nuclear-norm ball; the complexity is $\bigO(N( L_{\rm min}^2 L_{\rm max} + L_{\rm min} \log(L_{\rm min}) ))$, where $L_{\rm min} = \min\{L_x,L_y\}, L_{\rm max}= \max\{L_x,L_y\}$.

For the step-size selection, we first state the result.
If $\setS$ is the column-wise unit-simplex in \eqref{eq:setAS}, we choose
\begin{equation} \label{eq:beta_S_1}
\hat{\beta}_S^k =  \max\{ \hbetamin{S}, \lammax( (\bA^{k+1} \bPsi)^\top (\theta_G \bI + \bF^\top \bF) (\bA^{k+1} \bPsi) ) \},
\end{equation}
where $\theta_G = \lammax(\bG^\top \bG)$; $\bPsi \in \Rbb^{N \times (N-1)}$ is any semi-orthogonal matrix such that $\bPsi^\top \bone= \bzero$;
$\hbetamin{S} > 0$ is a small pre-fixed constant to safeguard $\beta_S^k$ from becoming too small.
If $\setS$ is the row-wise nuclear-norm ball in \eqref{eq:setS_NNB}, we choose
\begin{equation} \label{eq:beta_S_2}
\hat{\beta}_S^k =  \max\{ \hbetamin{S}, \lammax( (\bA^{k+1} )^\top (\theta_G \bI + \bF^\top \bF) (\bA^{k+1}) ) \}.
\end{equation}
As will be explained, the above rule is the ``best'' choice.
Eq.~\eqref{eq:beta_S_1} requires us to compute the largest eigenvalue of a PSD matrix of size  $(N-1) \times (N-1)$.
It can be verified that the complexity of \eqref{eq:beta_S_1} is  $\bigO(N^2 M + M^2 N + N^3) =\bigO(M^2 N ) $ (as $N \leq M$);
note that $\theta_G$ can be computed before the algorithm commences.
Similarly, the complexity of \eqref{eq:beta_S_2} is $\bigO(M^2 N)$.

Now we show the principle and derivations that lead to the step-size rule in \eqref{eq:beta_S_1}--\eqref{eq:beta_S_2}.
We simplify the notations by letting
\begin{equation} \label{eq:S_simplify}
\begin{gathered}
\bs^+ = \vvec(\bS^{k+1}), ~ \bs = \vvec(\bS^{k}), ~ \bs^- = \vvec(\bS^{k-1}),
\alpha = \alpha_k, ~ \bz = \vvec(\bS^{k}_{\sf ex}) = \bs + \alpha(\bs - \bs^-), \\
\beta = \hat{\beta}_S^k, ~ \bA = \bA^{k+1}, ~
 \tilde{\setS}= \{ \bs \in \vvec(\bS) \mid \bS \in \setS \},
f_s(\bs) = f(\bA,\bS) = \tfrac{1}{2} \| \by - \bH\bs \|^2,   \\
\by = \begin{bmatrix} \vvec(\bYm) \\ \vvec(\bYh) \end{bmatrix},
\quad
\bH = \begin{bmatrix} \bI_L \otimes (\bF \bA) \\ \bG^\top \otimes \bA \end{bmatrix}
\end{gathered}
\end{equation}
such that the FPG update in \eqref{eq:HiBCD_CoSMF_S} can be simplified to
\[
\bs^+ = \prox_{\tilde{\setS}}\left( \bz - \tfrac{1}{\beta} \nabla f_s(\bz) \right).
\]
We choose $\beta > 0$
such that the sufficient descent condition
\begin{equation} \label{eq:suff_dec_S}
f_s(\bs^+) \leq f_s(\bz) + \langle \nabla f_s(\bz), \bs^+ - \bz \rangle + \tfrac{\beta}{2} \| \bs^+ - \bz \|^2
\end{equation}
holds \cite{beck2017first}.
Also we want $\beta$ to be as small as possible,
{as this will make the step size $1/\beta$ as large as possible, and thereby the progress made at each FPG update maximized.}
We apply Fact~\ref{fac:quad_qub} to obtain such a $\beta$.
\begin{Fact} \label{fac:Lip_s_1}
    Let $\setS$ be the column-wise unit-simplex in \eqref{eq:setAS}.
    The smallest $\beta$ for which \eqref{eq:suff_dec_S} holds for any feasible $\bs^+$ and $\bz$ is $\beta = \lammax(\bR_S)$,  where
    \begin{equation} \label{eq:R_S}
    \bR_S = \bI_L \otimes [ (\bF \bA \bPsi )^\top (\bF \bA \bPsi ) ] + ( \bG \bG^\top ) \otimes [ (\bA \bPsi)^\top (\bA \bPsi) ],
    \end{equation}
    and $\bPsi \in \Rbb^{N \times (N-1)}$ is any semi-orthogonal matrix such that $\bPsi^\top \bone = \bzero$.
\end{Fact}

{\em Proof:} \
It is shown in \cite{lin2015identifiability} that ${\rm aff} \, \setU^N = \{ \bs = \bPsi \btheta +  \frac{1}{N} \bone \mid \btheta \in \Rbb^{N-1}  \}$.
Consequently, it can be verified that ${\rm aff} \, \tilde{\setS} = \{ \bs = \bPhi \bm \xi +  \frac{1}{N} \bone \mid \bm \xi \in \Rbb^{L(N-1)}  \}$, where $\bPhi = \bI_L \otimes \bPsi$;
and that the affine hull of the feasible set of $\bz$ is ${\rm aff} \, \tilde{\setS}$.
Invoking Fact~\ref{fac:quad_qub}, the smallest $\beta$ for which \eqref{eq:suff_dec_S} holds for any feasible $\bs^{+}$ and $\bz$ is $\beta = \lammax(\bPhi^\top \bH^\top \bH \bPhi)$.
As a routine exercise with Kronecker product, $\bPhi^\top \bH^\top \bH \bPhi$ is given by \eqref{eq:R_S}.
\hfill $\blacksquare$

\medskip

The matrix $\bR_S$ is $(N-1)L$-by-$(N-1)L$, and computing $\lammax(\bR_S)$ directly is expensive for large $L$.
The proposition below shows that $\lammax(\bR_S)$ can be obtained by computing the largest eigenvalue of an $(N-1) \times (N-1)$ matrix.
\begin{Prop} \label{prop:Lip_s_2}
    The largest eigenvalue of $\bR_S$ in \eqref{eq:R_S} equals
    \begin{equation} \label{eq:Lip_S_eff}
    \lammax(\bR_S) =
    \lammax( (\bA \bPsi)^\top (\theta_G \bI  + \bF^\top \bF) (\bA \bPsi) ),
    \end{equation}
    where $\theta_G = \lammax(\bG^\top \bG)$.
\end{Prop}

{\em Proof: } \
Let  $\bG \bG^\top = \bU \bLam \bU^\top$ be
 the eigendecomposition of $\bG \bG^\top$, where
 $\bU$ is orthogonal; $\bLam = \Diag(\blam)$; $\lambda_1 \geq \ldots \geq \lambda_L \geq 0$.
Let $\bQ = \bU^\top \otimes \bI_{N-1}$.
We have
\begin{align*}
\bQ^\top \bR_S \bQ & = \bI_L \otimes [ (\bF \bA \bPsi )^\top (\bF \bA \bPsi ) ] + \bLam \otimes [ (\bA \bPsi)^\top (\bA \bPsi) ]
 = {\rm Blkdiag}(\bB_1,\ldots,\bB_L),
\end{align*}
where ${\rm Blkdiag}$ denotes the block diagonal version of ${\rm Diag}$;
$\bB_i = (\bF \bA \bPsi )^\top (\bF \bA \bPsi ) + \lambda_i (\bA \bPsi)^\top (\bA \bPsi)$.
Since $\bQ$ is orthogonal, we are led to
\begin{align*}
\lammax( \bR_S ) & = \lammax( \bQ^\top \bR_S \bQ )
= \max_{i=1,\ldots,L} \lammax(\bB_i)
= \lammax(\bB_1),
\end{align*}
where the second equality is due to the block diagonal structure of $\bQ^\top \bR_S \bQ$;
the third inequality is due to the facts that  $\bB_1 \succeq \bB_i$ for all $i$ and
 that  $\bX \succeq \bY \Longrightarrow \lammax(\bX) \geq \lammax(\bY)$
 {(see, e.g., \cite[(4.13)]{calafiore2014optimization})}.
Also,
by noting $\lambda_1 = \lammax(\bG\bG^\top ) = \lammax(\bG^\top \bG)$, we obtain the desired result.
\hfill $\blacksquare$
\medskip

Eq.~\eqref{eq:Lip_S_eff} leads us to the step-size rule \eqref{eq:beta_S_1} for column-wise unit-simplex $\setS$.
The step-size rule \eqref{eq:beta_S_2} for row-wise nuclear-norm ball $\setS$ is shown by the same way, with $\bPsi = \bI$.

The FPG update of $\bA$ in \eqref{eq:HiBCD_CoSMF_A} follows the same development as above, and for conciseness we shall only state the results.
The gradient $\nabla_\bA f$ is
\[
\nabla_\bA f(\bA,\bS)  = \bF^\top (\bF \bA \bS - \bYm ) \bS^\top + (\bA\bS\bG - \bYh)(\bS \bG)^\top,
\]
and it can be computed in $\bigO(N(LM+\sum_{i=1}^{L_i}|\mathcal{L}_i|))
$ operations.
We have $\prox_\setA(\bZ) = [ \bZ ]_\bzero^\bone$.
The step-size rule is
\[
\hat{\beta}_A^k = \max\{ \hbetamin{A},  \lammax(\theta_F \bS^k (\bS^k)^\top + (\bS^k\bG)(\bS^k\bG)^\top ) \},
\]
where  $\hbetamin{A} > 0$ is a small pre-fixed constant; $\theta_F = \lammax(\bF \bF^\top)$.
We can compute $\hat{\beta}_A^k$ in {$\bigO(N^2 L)$} operations.

\subsection{The FW Updates}\label{sec:fw}

Next, we turn to the FW updates.
Consider the FW update of $\bS$ in \eqref{eq:HiBCD_CoSMF_S}.
We already provided the gradient formula in \eqref{eq:grad_S} and discussed its complexity.
Like the $\prox_\setS$ operations in the last subsection,
the $\LO_\setS$ operations are straightforward adaptations of the LO oracles in Section \ref{sec:prox_n_LO}.
For the case of column-wise unit-simplex $\setS$ in \eqref{eq:setAS}, $\LO_\setS$ does not incur floating-point operations.
In comparison,
$\prox_\setS$ for the same $\setS$ takes $\bigO(LN\log(N))$ operations.
It is also worth noting that $\LO_\setS$ for the case of row-wise nuclear-norm ball $\setS$ in \eqref{eq:setS_NNB} requires us to solve a number of $N$ principal singular vector problems,
while $\prox_\setS$ for the same $\setS$ requires $N$ full SVDs and $N$ unit-simplex projections.
We employ the power method to deal with the principal singular vectors problems.
In particular, we apply a warm-start trick wherein we use the principal singular vectors in the last iterate $k$ as the starting point of the power method in the present iterate $k+1$.

For the step size, our chosen rule is
\begin{equation} \label{eq:FW_S_stepsize}
\gamma_{S}^k =
1 \wedge \frac{ - \langle \nabla_\bS f(\bA^{k+1},\bS^k), \bD_S^k \rangle }{\| \bA^{k+1} \bD_S^k \bG \|_F^2 + \| \bF \bA^{k+1} \bD_S^k  \|_F^2  + \hbetamin{S} \| \bD_S^k \|_F^2 },
\end{equation}
where $\bD_S^k = \LO_\setS(\nabla_\bS f(\bA^{k+1},\bS^k )) - \bS^k$; $\hbetamin{S}$ is a small pre-fixed constant;
and recall $a \wedge b = \min\{ a, b \}$.
The complexity of \eqref{eq:FW_S_stepsize} is $\bigO(N(ML+\sum_{i=1}^{L_H}|\mathcal{L}_i|))$.
The step-size rule \eqref{eq:FW_S_stepsize} is a variant of the adaptive step-size rule in \cite{beck2015cyclic}.
We begin by describing the latter.
By adopting the simplified notations in \eqref{eq:S_simplify}, and additionally, $\gamma = \gamma_{S}^k, \bD = \bD_S^k, \bd = \vvec(\bD)$, we simplify the FW update of $\bS$ in \eqref{eq:HiBCD_CoSMF_S} to $\bs^+ = \bs + \gamma \bd$.
Consider the sufficient descent condition
\begin{align}
f_s(\bs^+) & \leq f_s(\bs) + \langle \nabla f_s(\bs), \bs^+ - \bs \rangle + \tfrac{\beta}{2} \| \bs^+ - \bs \|^2
= f_s(\bs) +  \gamma \langle \nabla f_s(\bs), \bd \rangle + \tfrac{\gamma^2 \beta}{2} \| \bd \|^2,
\label{eq:FW_qup_orig}
\end{align}
for some $\beta > 0$, which, as studied previously, can be achieved by choosing $\beta$ as \eqref{eq:beta_S_1} or \eqref{eq:beta_S_2}.
We choose $\gamma$ by minimizing the quadratic upper bound in \eqref{eq:FW_qup_orig} over $[0,1]$, i.e.,
\begin{align}
\gamma & = \arg \min_{\bar{\gamma} \in [0,1]} \bar{\gamma} \langle \nabla f_s(\bs), \bd \rangle + \tfrac{ \bar{\gamma}^2 \beta}{2} \| \bd \|^2
= 1 \wedge  \frac{-\langle  \nabla f_s(\bs), \bd \rangle}{\beta \| \bd \|^2 }.
\label{eq:FW_stepsize_org}
\end{align}
Our step-size rule variant considers
\begin{align}
f_s(\bs^+) & \leq f_s(\bs) + \langle \nabla f_s(\bs), \bs^+ - \bs \rangle + \tfrac{1}{2} \| \hat{\bR}^{1/2} (\bs^+ - \bs) \|^2
\nonumber \\
& = f_s(\bs) +  \gamma \langle \nabla f_s(\bs), \bd \rangle + \tfrac{\gamma^2}{2} \| \hat{\bR}^{1/2} \bd \|^2,
\label{eq:FW_qup_us}
\end{align}
for some symmetric positive definite (PD) $\hat{\bR}$; here $\hat{\bR}^{1/2}$ denotes the PSD square root of $\hat{\bR}$.
By the quadratic structure of $f_s$, it is easy to show that a choice of $\hat{\bR}$ for making \eqref{eq:FW_qup_us} happen is $\hat{\bR} = \bH^\top \bH + {\hbetamin{}} \bI $ for any ${\hbetamin{}} > 0$;
in fact, equality in \eqref{eq:FW_qup_us} approaches zero as $\delta \rightarrow 0$.
By choosing $\gamma$ as the minimizer of the quadratic upper bound \eqref{eq:FW_qup_us} over $[0,1]$, we obtain the step-size rule
\begin{align} \label{eq:FW_S_stepsize_abs}
\gamma & =
1 \wedge \frac{-\langle \nabla f_s(\bs), \bd \rangle}{ \| \hat{\bR}^{1/2} \bd \|^2 }.
\end{align}
Finally, by putting $\| \hat{\bR}^{1/2} \bd \|^2= \| \bA\bD\bG\|_F^2 + \| \bF\bA\bD \|_F^2 + {\hbetamin{}}  \| \bD \|_F^2$ into \eqref{eq:FW_S_stepsize_abs}, the step-size rule in \eqref{eq:FW_S_stepsize} is yielded.

The FW update of $\bA$ in \eqref{eq:HiBCD_CoSMF_A} follows the same development as above, and  for brevity we shall omit the details.

%

\section{Analysis of HiBCD}
\label{sec:analysis_hibcd}

The proposed HiBCD scheme can be regarded as an instance of a more general optimization technique.
To put into context, let
$\bx_i \in \Rbb^{n_i}$,
$i=1,\ldots,m$,
and let $\bx =  (\bx_1,\ldots,\bx_m) \in \Rbb^n$ where $\sum_{i=1}^m n_i = n$. Consider
\beq
\label{eq:Pmain}
\begin{aligned}
    \min_{\bx \in \Rbb^n } & ~ F(\bx) := f(\bx) + h(\bx),
\end{aligned}
\eeq
where $f: \Rbb^n \rightarrow \Rbb$ is differentiable and can be non-convex;
$h(\bx)$ takes the form
\[ \textstyle
h(\bx) = \sum_{i=1}^m h_i(\bx_i),
\]
in which every $h_i: \Rbb^{n_i} \rightarrow (-\infty,\infty]$ is convex, closed, proper, and possibly non-smooth;
every domain
\[
\setX_i : = {\rm dom} ( h_i )
\]
is convex and compact.
For notational convenience,
we will denote
$\setX = \setX_1 \times \cdots \times \setX_m$,
$\bx_{-i}  = ( \bx_j )_{j=1, j \neq i}^m$,
$\setX_{-i}$ as the domain of $\bx_{-i}$,
$f(\bx_i,\bx_{-i})$ as an alternative of writing $f(\bx)$,
and $\nabla_if(\bx) = \nabla_{\bx_i} f(\bx)$.

Using the idea of hybrid coordinate descent mentioned before, the HiBCD scheme assigns one of the two update rules (FPG or FW) to each variable $\bx_i$. We use $\setI_{\sf FPG}$ (\resp $\setI_{\sf FW}$) to denote the set of variables updated by FPG (\resp FW). We summarize the HiBCD scheme for \eqref{eq:Pmain} in Algorithm~\ref{alg:hibcd}. Note that the scheme
supports repeating the updates for multiple ($L_i$) times within the same block.
We remark that the step sizes for FPG/FW are selected
such that the sufficient descent conditions
in \eqref{eq:fpg_step_full}--\eqref{eq:fw_step_full} are satisfied;
they are standard and apply to common step-size rules such as the backtracking line search and explicit Lipschitz constant rules \cite{beck2017first}.

{
\begin{algorithm}[!t]
    \caption{Hybrid Inexact BCD Scheme for \eqref{eq:Pmain}} \label{alg:hibcd}
    \begin{algorithmic}[1]
        \STATE \textbf{given:} a starting point $\bx^0$.
        \STATE $L_0=0$, $\bx^{0,0,0} = \bx^0$, $\bx^{0,0,-1} = \bx^0$ \COMMENT{initialize}
        \FOR {$k= 0, 1,2,\ldots,K$}
        \FOR {$i=1,2, \dots, m$}
        \STATE $\bx^{k,i,0} = \bx^{k,i-1,L_{i-1}}$
        \STATE $\bx^{k,i,-1} = \bx^{k,i-1,L_{i-1}-1}$ \COMMENT{previous round}
        \FOR {$\ell=0,...,L_i-1$}
        \IF{$i \in \setI_{\sf FPG}$}\label{line:hibcd1}
        \STATE set $\bx_{i,{\sf ex}}^{k,i,\ell} = \bx_{i}^{k,i,\ell} + \alpha_{i}^{k,\ell} \big( \bx_{i}^{k,i,\ell}  - \bx_{i}^{k,i,\ell-1}  \big)$ \COMMENT{extrapolation}
        \STATE Perform the FPG update $$\bx_{i}^+ =
        \prox_{h_i / \hat\beta_i^{k,\ell}}
        \big( \bx_{i,{\sf ex}}^{k,i,\ell} - {\textstyle \frac{1}{ \hat\beta_i^{k,\ell} }} \grd_i f( \bx_{i,{\sf ex}}^{k,i,\ell}, \bx_{-i}^{k,i,\ell} ) \big)$$
        where $\hat{\beta}_i^{k,\ell}$ is chosen such that
        \beq \label{eq:fpg_step_full}
        f( \bx_i^+ \!, \bx_{-i}^{k,i,\ell} ) \!\leq\! f( \bx_{i,{\sf ex}}^{k,i,\ell} , \bx_{-i}^{k,i,\ell} ) + \langle \grd_i f(\bx_{{\sf ex}}^{k,i,\ell} ),  \bx_i^+ - \bx_{i,{\sf ex}}^{k,i,\ell} \rangle  + \frac{ \hat{\beta}_i^{k,\ell}}{2} \| \bx_i^+ - \bx_{i,{\sf ex}}^{k,i,\ell} \|^2
        \eeq

        \ENDIF

        \IF{$i \in \setI_{\sf FW}$}
        \STATE Perform the FW update
        $$\bx_i^+ = \bx_{i}^{k,i,\ell} + \gamma_i^{k,\ell} \bd_i^{k,i,\ell}$$
        where $\by_i^{k,i,\ell} = \LO_{h_i} (\grd_i f( \bx^{k,i,\ell} ) )$, $\bd_i^{k,i,\ell} = \bx_i^{k,i,\ell} - \by_i^{k,i,\ell}$,
        \beq \notag
        \gamma_i^{k,\ell} = 1 \wedge \frac{ \langle \grd_i f ( \bx^{k,i,\ell} ), \bd_i^{k,i,\ell} \rangle + h_i (\bx_i^{k,i,\ell}) - h_i(\by_i^{k,i,\ell}) }{ \| \hat{\bm R}_{k,i,\ell}^\frac{1}{2} \bd_i^{k,i,\ell}  \|^2 },
        \eeq
        and $\hat{\bR}_{k,i,\ell}$ is chosen such that it is PD and satisfies
        \beq \label{eq:fw_step_full}
        f( \bx_i^+, \bx_{-i}^{k,i,\ell} ) \leq f( \bx_{i}^{k,i,\ell} , \bx_{-i}^{k,i,\ell} ) + \langle \grd_i f(\bx^{k,i,\ell} ),  \bx_i^+ - \bx_{i}^{k,i,\ell} \rangle  + \frac{1}{2} \| \hat{\bR}_{k,i,\ell}^{\frac{1}{2}} (\bx_i^+ - \bx_{i}^{k,i,\ell}) \|^2
        \eeq

        \ENDIF

        \STATE Set $\bx_{i}^{k,i,\ell+1} = \bx_{i}^+$, $\bx_{-i}^{k,i,\ell+1} = \bx_{-i}^{k,i,\ell}$ \label{line:hibcd3} 
        \ENDFOR
        \ENDFOR
        \STATE $\bx^{k+1} = \bx^{k,m,L_m}$
        \STATE $\bx^{k+1,0,0} = \bx^{k,m,L_m}, \bx^{k+1,0,-1}= \bx^{k,m,L_m-1}$
        \ENDFOR
    \end{algorithmic}
\end{algorithm}}

Our
interest lies in showing sufficient conditions under which the HiBCD scheme guarantees some form of convergence to a stationary point to Problem~\eqref{eq:Pmain}.
A point $\hat{\bx} \in \setX$ is said to be a stationary point to Problem~\eqref{eq:Pmain} if $- \nabla f(\hat{\bx}) \in \partial h(\hat{\bx})$ \cite{bertsekas2003convex}.
The stationarity condition can be verified by the
FW gap
\begin{equation} \label{eq:fw_gap_def} \textstyle
g(\bx) = 
\max_{\by \in \RR^n } \langle \nabla f(\bx), \bx - \by \rangle + h(\bx) - h(\by).
\end{equation}
In particular, a point $\bx \in \setX$ attains $g(\bx) = 0$ if and only if it is a stationary point to problem \eqref{eq:Pmain}  \cite[Theorem 13.6]{beck2017first}.
Also, it is true that $g(\bx) \geq 0$ for all $\bx \in \setX$.

To establish the convergence of HiBCD, we need to define an extension to the domain $\setX$ as the algorithm may encounter extrapolated iterates that do not reside in $\setX$:
\beq \notag
\tilde{\setX}_i = \hspace{-.1cm} \begin{cases}
    \ds {\rm conv} \big[
    \hspace{-.2cm}
    \bigcup_{ 0 \leq \alpha \leq 1 } \hspace{-.2cm} \big\{ \bx_i + \alpha( \bx_i - \by_i) \mid \bx_i,\by_i \in \setX_i \big\} \big], i \in \setI_{\sf FPG} \\
    \setX_i \hfill,~i \in \setI_{\sf FW}.
\end{cases}
\eeq
Let $\tilde\setX \eqdef \tilde\setX_1 \times \cdots \times \tilde\setX_m$, and assume the following.
\begin{Asm} \label{assumption}
    The function $f$ has $\beta$-Lipschitz continuous gradient on $\tilde\setX$.
\end{Asm}
As discussed in Section \ref{sec:Lip_cont}, Assumption \ref{assumption} implies that $f$ has block-wise Lipschitz continuous gradient on $\tilde\setX$.
To facilitate the description of the convergence result to be presented,
let $\beta_i$ be the block-wise tight Lipschitz constant of $\nabla_i f$ on $\tilde\setX$.
Also, define $\rho_i \in [0,\beta_i]$, $i=1,...,m$, such that $f$ is block-wise weakly convex on $\tilde\setX$ with parameters $\rho_1, \ldots, \rho_m $; as reviewed in Section \ref{sec:Lip_cont}, the weakly convex assumption is at least true for $\rho_i =\beta_i $, $i=1,...,m$.
Our convergence result is as follows.
\begin{Theorem} \label{thm2}
	Consider the HiBCD settings described above.
	Suppose that Assumption \ref{assumption}  holds;
	that there exists $\delta_i > 0, \eta_i \geq 1$ such that the step-size parameters in \eqref{eq:fpg_step_full}--\eqref{eq:fw_step_full} satisfy,
	\beq \label{eq:stepsize_cond}
	\delta_i \leq \hat{\beta}_i^{k,\ell} \leq \eta_i \beta_i ,~
	\delta_i {\bm I} \preceq \hat{\bm R}_{k,i,\ell} \preceq \eta_i \beta_i {\bm I},
	\eeq
	for all $k,\ell$; and
	that for every $i \in \setI_{\sf FPG}$, there exists $\bar{\aaalp{}} \in [0,1)$ such that
	\begin{equation}  \label{eq:alpha_cond}
	\alpha_{i}^{k,\ell} \leq \bar{\aaalp{}} \!~ \sqrt{ \frac{ \hat{\beta}_i^{k,\ell-1} }{ \rho_i + \hat{\beta}_{i}^{k,\ell}  } },
	\quad \forall k,\ell.
	\end{equation}
	Then the following results hold.
	\begin{enumerate}
		\item The convergence rate is sublinear; specifically,
		\beq \label{eq:sublinear}
		\min_{k=0,\ldots,K} g(\bx^k)^2 \leq \frac{m B C }{K+1} [ F(\bx^{0}) - F^\star  ],
		\eeq
		where $F^\star = \min_{\bx \in \mathbb{R}^n} F(\bx)$,
		\[
		B = 1 + \frac{ \bar{\aaalp{}}^2   }{ 1 - \bar{\aaalp{}}^2 }
		\left( 2 + {\max_{j \in \setI_{\sf FPG}}} \frac{\beta_j + \rho_j}{\hbetamin{j}} \right),
		\]
		and the constant $C = \max_{j=1,...,m} C_j$ where
		\beq \label{eq:constC}
		C_j = \frac{2A_j}{L_j} +  \frac{ 4 (M_j + l_j)^2 (L_j-1)+ 8 (D\beta)^2 L_j}{\hbetamin{j}},
		\eeq
		with
		\[
		\hspace{-.6cm} A_i = \begin{cases}
		2 \max\{ \eta_i \beta_i D_i^2, (M_i + l_i) D_i \}, &\hspace{-.2cm} i \in \setI_{\sf FW}, \\
		2 \max\{ [ (2\eta_i + 1) \beta_i + \rho_i ] D_i^2, (M_i + l_i) D_i \}, &\hspace{-.2cm}  i \in \setI_{\sf FPG},
		\end{cases}
		\]
		where $l_i$ is the tight Lipschitz constant of $h_i$ on $\setX_i$; $D_i = \max_{\bx_i,\by_i \in \setX_i} \| \bx_i - \by_i \|$; $M_i = \max_{ \bx \in \setX_i } \| \grd_i f(\bx) \|$; $D = \sqrt{\sum_{i=1}^m D_i^2}$
		\item Any limit point of $\{ \bx^k \}_{k \geq 0}$ is a stationary point to Problem \eqref{eq:Pmain}.
	\end{enumerate}
\end{Theorem}
For Theorem~\ref{thm2} to hold, we require several conditions on the step sizes and extrapolation weights.
First, \eqref{eq:stepsize_cond} requires that the step size parameters of the FPG/FW updates to be  positive and upper bounded in proportion to the block-wise Lipschitz constant $\beta_i$.
This can be guaranteed when the step sizes are chosen with certain rules such as the backtracking line search and explicit Lipschitz constant rules.
Here, the parameter $\eta_i \geq 1$ quantifies the quality of step size selection;
$\eta_i = 1$ is the best, while large $\eta_i$ refers to cases where the chosen step size is smaller than that required by theory.
Also, $\delta_i >0$ is to prevent the algorithm from getting into pathological cases that can cause divergence.
Second, \eqref{eq:alpha_cond} specifies an upper bound on the extrapolation weights. To understand the condition further, consider a simplified step size selection $\hat{\beta}_i^{k,\ell} = \beta_i$ for all $k,\ell$.
We observe that the constraint on the extrapolation weights depends on the weak convexity parameter $\rho_i$---if $\rho_i = 0$ (each block is convex), then $\alpha_i^{k,\ell}$ can be chosen to be
close to $1$; if $\rho_i = \beta_i$, then $\alpha_i^{k,\ell}$  can only be chosen to be as large as $\sqrt{1/2}$.



Let us use the CoSMF problem \eqref{eq:CoSMF} as an example. We have $h_1= \indfn{\setA}, h_2= \indfn{\setS}$.
As $f$ is twice differentiable, and $\setA$ and $\setS$ are compact, Assumption~\ref{assumption} is satisfied.
As $f$ is convex in $\bA$ (resp. $\bS$) given $\bS$ (resp. $\bA$), $f$ has $\rho_1 = \rho_2 = 0$.
The HiBCD scheme we design for CoSMF in Section~\ref{sec:cosmf} has the step-size rules satisfying $\eta_i = 1$ for the FPG case, and $\eta_i = (1 + \hbetamin{i}/\beta_i) \approx 1$ for the FW case.
In addition, the scheme is Algorithm \ref{alg:hibcd} with $L_1 = L_2 = 1$, i.e.,  one-time FPG/FW update per block.
If we choose large $L_1, L_2$, it becomes a ``quasi-exact'' BCD wherein FPG/FW updates are applied many times per block to yield nearly exact BCD updates.
Our stationarity result in Theorem~\ref{thm2} also cover this case; i.e.,  the quasi-exact BCD also guarantees (subsequence) convergence to a stationary point.



\begin{Remark}
	Our convergence analysis of HiBCD unifies those of the alternating proximal gradient method \cite{xu2013block,xu2017globally} and the CBCG method \cite{beck2015cyclic}, which are inexact BCDs with only FPG or FW updates, resp. Specifically, the convergence analyses in the above two works are different and incompatible.
	We adopt the FW gap analysis approach in CBCG, and the main challenge is to incorporate the FPG update into the FW gap analysis. In that regard, Lemma~\ref{lem:EPG_suff_dec} to be presented in the proof of Theorem~\ref{thm2}, which addresses the aforementioned challenge, is particularly important.
	We should mention that Lemma~\ref{lem:EPG_suff_dec} is reminiscent of \cite[Lemma~2.6]{beck2018primal} when no extrapolation is involved.
	Another salient feature is that our analysis covers multiple FPG/FW updates per block, while those of the aforementioned works considered one-time update.
\end{Remark}

\begin{Remark}
	It is also interesting to draw insights from the convergence analysis,
	specifically, what are the best numbers $L_i$'s such that the convergence speed predicted by \eqref{eq:sublinear} is the fastest.
	The question is identical to finding the $L_j$'s such that the constants $C_j$'s in \eqref{eq:constC} are the smallest.
	We have:
	\begin{Fact} \label{fac:C_opt}
		Consider the constant $C_j$ in \eqref{eq:constC}.
		Let $L_j^\star$ be an optimal $L_j$ that gives the smallest $C_j$.
			It must be true that
			\beq \label{eq:optLj}
			L_j^\star \leq
			\begin{cases}
				\lceil \sqrt{\eta_j/2} \rceil, & j\in\setI_{\sf FW} \\
				\big\lceil \sqrt{ (2\eta_j + 1 + \frac{\rho_j}{\beta_j} ) / 2} \big\rceil, & j \in \setI_{\sf FPG}
			\end{cases}
			\eeq
			Also, we have $L_j^\star = 1$ if $j \in \setI_{\sf FW}$ and $\eta_j \leq 4$, or if $j \in \setI_{\sf FPG}$, $\rho_j=0$, and $\eta_j \leq \frac{3}{2}$.
	\end{Fact}
	{The proof of Fact~\ref{fac:C_opt} is relegated to
    Appendix~\ref{app:fac:C_opt}.}
	Fact~\ref{fac:C_opt} indicates that, under a good step-size selection such that $\eta_i$ is close to $1$, there may be not much benefit with applying multiple FPG/FW updates per block.
\end{Remark}


\subsection{Proof of Theorem~\ref{thm2}} \label{sec:analysis}
Our proof is divided into two steps. The first step quantifies the progress made in
one FPG/FW update.
The second step combines these result to prove convergence.\vspace{.2cm}

\noindent \textbf{Step 1: One-step Progress}. The first step in our proof is to derive  descent lemmas for FPG/FW updates, i.e., line~\ref{line:hibcd1}--\ref{line:hibcd3} in Algorithm~\ref{alg:hibcd}. We focus on bounding the progress made in one update. As such, we shall use the simplified  notations
\begin{gather*}
\bx^+ = \bx^{k,i,\ell+1},~\bx = \bx^{k,i,\ell},~\bx^- = \bx^{k,i,\ell-1}, \\
\bz = (\bx^{k,i,\ell}_1, \ldots, \bx^{k,i,\ell}_{i-1}, \bx_{i,{\sf ex}}^{k,i,\ell}, \bx^{k,i,\ell}_{i+1},\ldots,\bx^{k,i,\ell}_m ),
\end{gather*}
only in this paragraph.
Similarly we drop the superscripts of $k,\ell$ for
$\alpha_i^{k,\ell}, \hat\beta_i^{k,\ell}, \gamma_i^{k,\ell}$. Define the $i$th block's FW gap as
\begin{equation} \label{eq:fw_gapi_def}
g_i(\bx) = \max_{\by_i \in \RR^{n_i} } ~\big\{ \langle \nabla_i f(\bx), \bx_i - \by_i \rangle + h_i(\bx_i) - h(\by_i) \big\}.
\end{equation}

The one-step progress made by the FPG/FW updates can be summarized as follows. First, we focus on the FPG update:
\begin{Lemma} \label{lem:EPG_suff_dec}
	For $i \in \setI_{\sf FPG}$,
	\begin{gather}
	\label{eq:EPG_suff_dec1}
	g_i(\bx)^2 \leq
	\tilde{A}_i
	\left( F(\bx)- F(\bx^+) + \frac{ \bar{\beta}_i \aalp^2}{2} \| \bx_i - \bx_i^- \|^2 \right), \\
	\label{eq:EPG_suff_dec2}
	\| \bx_i^+ - \bx_i \|^2 - \big(1+ \frac{\rho_i}{\hat{\beta}_i} \big) \aalp^2 \| \bx_i - \bx_i^- \|^2  \leq \frac{2}{\hbetai} (F(\bx)- F(\bx^+)),
	\end{gather}
	where $\tilde{A}_i = 2  \max\{ \bar{\beta}_i D_i^2, (M_i + l_i) D_i \}$, $\bar{\beta}_i = 2 \hbetai + \beta_i + \rho_i$.
\end{Lemma}
The proof can be found in Appendix~\ref{app:fpg}.
For the FW update, we have the following lemma:
\begin{Lemma}  \label{lem:FW2_suff_dec}
	For $i \in \setI_{\sf FW}$,
	\begin{align}
	g_i(\bx)^2 & \leq
	\tilde{A}_i
	(F(\bx)- F(\bx^+)), \notag\\
	\| \bx_i^+ - \bx_i  \|^2 & \leq \frac{2}{ \delta_i } (F(\bx)- F(\bx^+)), \notag
	\end{align}
	where $\tilde{A}_i = 2 \max\{ \eta_i \beta_i D_i^2, (M_i + l_i) D_i \}$.
\end{Lemma}
This lemma is a straightforward extension of \cite[Lemma 4.6-4.7]{beck2015cyclic} and its proof is omitted.\vspace{.2cm}

\noindent \textbf{Step 2: Combining the Lemmas}.
We apply the results in Step 1 to derive a bound for $g( \bx^k )$.
To facilitate this,
define
\begin{align*}
\dF{k,i,\ell} & \eqdef F(\bxkil) - F(\bxki{\ell+1}), \\
\dx{k,i,\ell} & \eqdef \| \bxkil - \bxki{\ell+1} \|^2.
\end{align*}
The following bounds are direct summaries of
Lemmas~\ref{lem:EPG_suff_dec}--\ref{lem:FW2_suff_dec}. For any $i \in \{1,\ldots,m\}$,
\begin{align}
g_i(\bxkil)^2 & \leq
A_i \left(  \dF{k,i,\ell} + \frac{\bar{\beta}_{i}^{k,\ell} (\alpha_i^{k,\ell})^2 }{2} \dx{k,i,\ell} \right)
\label{eq:thm1_step1_eq1}
\\
\dx{k,i,\ell} & \leq \frac{2}{\hat{\beta}_{i}^{k,\ell}} \dF{k,i,\ell} + \big(1+ \frac{\rho_i}{\hat{\beta}_i^{k,\ell}} \big) (\alpha_i^{k,\ell})^2 \dx{k,i,\ell-1},
\label{eq:thm1_step1_eq2}
\end{align}
where we have defined $A_i$ after \eqref{eq:constC}; $\bar{\beta}_{i}^{k,\ell} = 2 \hat{\beta}_{i}^{k,\ell} +\beta_i + \rho_i$ for all $i \in \{1,...,m\}$;
if $i \in \setI_{\sf FW}$,
we define $\alpha_i^{k,\ell} = 0$, $\hat{\beta}_{i}^{k,\ell} = \hbetamini$ and $\bar{\beta}_{i}^{k,\ell} = 0$.
Furthermore, we shall work with the quantities
\beq
\dF{k,j} \eqdef \sum_{\ell=0}^{L_j-1} \dF{k,j,\ell},~
\tdx{k,j} \eqdef \sum_{\ell=0}^{L_j-1} \frac{\bar{\beta}_{j}^{k,\ell} (\alpha_j^{k,\ell})^2}{2} \dx{k,j,\ell-1}.\notag
\eeq
We first observe the following lemma:
\begin{Lemma}{\cite{beck2015cyclic}}
	\label{fact:Lip_g} It holds that for any $\bx, \by \in \setX$,
	\[
	| g_i(\bx) - g_i(\by)| \leq D_i \beta \| \bx - \by \| + (M_i + l_i ) \| \bx_i - \by_i \|.
	\]
\end{Lemma}
Lemma~\ref{fact:Lip_g} indicates that $g_i$ is Lipschitz continuous on $\setX$.
It is an extension of \cite[Lemma 3.2]{beck2015cyclic} and we omit the proof.
Lemma~\ref{fact:Lip_g}, together with \eqref{eq:thm1_step1_eq1}--\eqref{eq:thm1_step1_eq2},
lead to the result below:
\begin{Lemma} \label{lem:thm_step2}
	For any $i \in \{1,...,m\}$, it holds that
	\begin{equation} \label{eq:thm_step2_main_eq}
	g_i(\bx^k)^2 \leq \sum_{j=1}^m \bar{C}_{ij} ( \dF{k,j} + \tdx{k,j} ),~~~~~~\text{where}
	\end{equation}
	\[
	\bar{C}_{ij}  = \begin{cases}
	\displaystyle \frac{8(D_i \beta)^2 L_j}{\hbetamin{j}},~1 \leq j \leq i-1 \\
	\displaystyle \frac{2A_i}{L_i} + \frac{[ (D_i \beta)^2 + (M_i + l_i)^2 ] (L_i - 1)}{\hbetamini / 4},~i=j, \\
	0,~~~\text{otherwise}
	\end{cases}
	\]
\end{Lemma}
{The proof  is shown in
Appendix~\ref{app:lemstep2}.}

We observe that $g(\bx) = \sum_{i=1}^m g_i(\bx)$; cf. \eqref{eq:fw_gap_def}, \eqref{eq:fw_gapi_def}.
Thus, 
\begin{align}
& \frac{1}{m} g(\bx^k)^2 \leq \sum_{i=1}^m g_i(\bx^k)^2
 \leq  \sum_{i=1}^m \sum_{j=1}^m \bar{C}_{ij} ( \dF{k,j} + \tdx{k,j} ) \leq \bar{C} \sum_{j=1}^m ( \dF{k,j} + \tdx{k,j} )
 \notag
\end{align}
where $\bar{C}= \max_{j=1,\ldots,m} \bar{C}_j$, $\bar{C}_j = \sum_{i=1}^m \bar{C}_{ij}$.
Notice that $C_j$ in \eqref{eq:constC} satisfies $\bar{C}_j \leq C_j$.
Also,
\begin{align}
& \sum_{j=1}^m \dF{k,j} = \sum_{j=1}^m  ( F(\bxk{j}{0}) -  F(\bxk{j}{L_j}) )
= F(\bxk{1}{0}) -  F(\bxk{m}{L_m})  = F(\bx^k) - F(\bx^{k+1}),
\notag
\end{align}
where we have used $\bxk{j}{L_j} = \bxk{j+1}{0}$.
We thus obtain
\beq \label{eq:gsum}
g(\bx^k)^2 \leq mC \Big( F(\bx^k) - F(\bx^{k+1}) + \sum_{j=1}^m  \tdx{k,j} \Big),
\eeq
where $C = \max_{j=1,\ldots,m} C_j$.
Summing both sides of \eqref{eq:gsum} from $k=0$ to $k=K$ yields
\beq \notag \textstyle
\sum_{k=0}^K g(\bx^k)^2 \leq mC \Big( F(\bx^0) - F^\star + \sum_{k=0}^K \sum_{j=1}^m  \tdx{k,j} \Big),
\eeq
where we have used 
$F( \bx ) \geq F^\star$ for any $\bx \in \setX$.
Since the extrapolation weights satisfy \eqref{eq:alpha_cond}, it can be shown that
\beq \label{eq:fpgtelescope}
\sum_{k=0}^K \sum_{i=1}^m \tdx{k,i} \leq   \left(2 + {\max_{j \in \setI_{\sf FPG}}} \frac{\beta_j + \rho_j}{\delta_j} \right) \frac{\bar{\alpha}^2}{1-\bar{\alpha}^2}  \sum_{k=0}^K \sum_{i=1}^m \dF{k,i}
\eeq
{The proof of \eqref{eq:fpgtelescope} is technical, and we relegate it to
Appendix~\ref{app:fpgtele}.}
Combining
the above derivations
gives
\beq \notag \textstyle
\sum_{k=0}^K g(\bx^k)^2 \leq {mBC} \left( F(\bx^0) - F^\star  \right).
\eeq
Lower bounding the left-hand side of the above inequality by $(K+1) \min_{k=0,...,K} g(\bx^k)^2$ concludes the proof of the first statement in Theorem~\ref{thm2}.

To prove
the second statement of Theorem~\ref{thm2}, we begin by defining $\tdx{k} \eqdef \sum_{i=1}^m \tdx{k,i}$. Observe from \eqref{eq:fpgtelescope} that
$\lim_{k \rightarrow \infty} \tdx{k} = 0$ as the right-hand side of \eqref{eq:fpgtelescope} is finite. Consider a subsequence $\{ \bx^{k_\ell} \}_{\ell \geq 0}$ with limit $\bar{\bx}$. Note that the limit $\bar{\bx}$ exists as $\setX$ is compact. From \eqref{eq:gsum}, we observe
\beq
\begin{split}
	g(\bx^{k_\ell})^2 & \textstyle \leq \sum_{k=k_\ell}^{k_{\ell+1}-1} g(\bx^{k_\ell})^2
    \leq mC \Big[ F(\bx^{k_\ell}) - F(\bx^{k_{\ell+1}} ) + {\textstyle \sum_{k=k_\ell}^{k_{\ell+1}-1} \tdx{k} } \Big]
    \notag
\end{split}
\eeq
Define
the non-negative number
$s_\ell \eqdef mC [ F(\bx^{k_\ell}) - F(\bx^{k_{\ell+1}} ) + {\textstyle \sum_{k=k_\ell}^{k_{\ell+1}-1} \tdx{k} } ]$. We note that $s_\ell \rightarrow mC [ F(\bar{\bx}) - F(\bar{\bx} ) ] = 0$.
It follows from the continuity of $g(\cdot)$ (cf.~Lemma~\ref{fact:Lip_g}) that $\bar{\bx}$ satisfies $g(\bar{\bx}) = 0$ . 


\section{Numerical Experiments}
\label{sec:num_exp}

\subsection{Semi-Real Experiment}

We follow a standard procedure, namely, Wald's protocol \cite{WALDS_PROTOCOL,loncan2015hyperspectral}, to perform  a semi-real experiment.
The ground-truth SR image $\bX$ is a real image from the Hyperspec Chikusei dataset \cite{NYOKOYA2016}.
The image size is $(L_x,L_y) = (1,080,1,080)$, and the number of spectral bands is $M= 128$.
The MS-HS image pair $(\bYm,\bYh)$ is generated by the model in \eqref{eq:Ym_model} and \eqref{eq:Yh_model}.
The spectral decimation follows the specification of the IKONOS MS sensor  \cite{dial2003ikonos}, with the number of MS spectral bands given by $\Mm= 4$.
The spatial decimation corresponds to $11 \times 11$ truncated Gaussian spreading with variance $1.7^2$, followed by downsampling with a factor of $8$;
the resulting number of HS pixels is $\Lh= 135^2$.
The noise terms $\bVm$ and $\bVh$ are randomly generated, following an i.i.d. mean-zero Gaussian distribution.
The SNR is $20$dB.
This problem is considered large;  we have $L= 1,166,400$.

We benchmark the proposed HiBCD scheme with CNMF \cite{yokoya2012coupled}, SupResPALM \cite{lanaras2015hyperspectral}, and FUMI \cite{wei2016multiband}.
All the algorithms have the same model order, $N=20$, and the same initialization.
Our initialization follows that in \cite{lanaras2015hyperspectral}.
The stopping rule of the algorithms is that either the objective value change is lower than $10^{-4}$, or the iteration number reaches $3,000$.
We also consider a {\em naive interpolation} baseline where we apply bicubic interpolation to each spectral band of the HS image to produce an SR image.
Unless specified, the HiBCD scheme employs the FPG update for $\bA$ and the FW update for $\bS$.
{The FISTA extrapolation sequence in \eqref{eq:fista_seq} is employed for the FPG update.}
For NNC CoSMF, we set $\tau= 300$ for all $i$.

\begin{figure}[!tbt]
	\begin{minipage}[c]{\linewidth}
		\qquad\qquad\includegraphics[width=.8\linewidth]{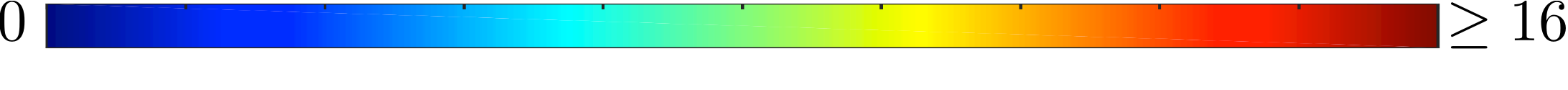}
		\begin{minipage}[b]{.33\textwidth}
			\centering
			\includegraphics[width=\linewidth]{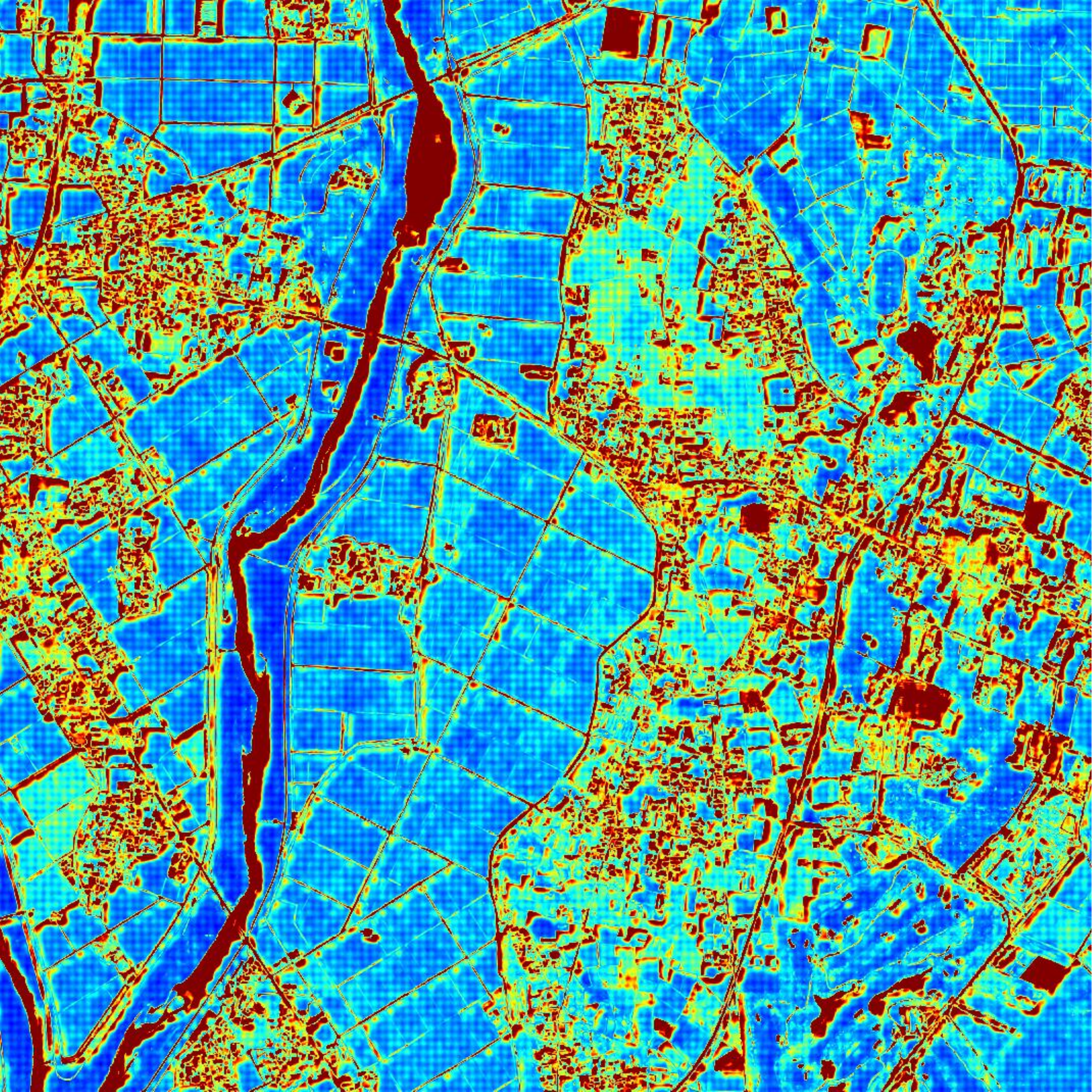}
			(a) naive interpolation\\
			(mean SAM $=8.85^\circ$)
		\end{minipage}
		\begin{minipage}[b]{.33\textwidth}
			\centering
			\includegraphics[width=\linewidth]{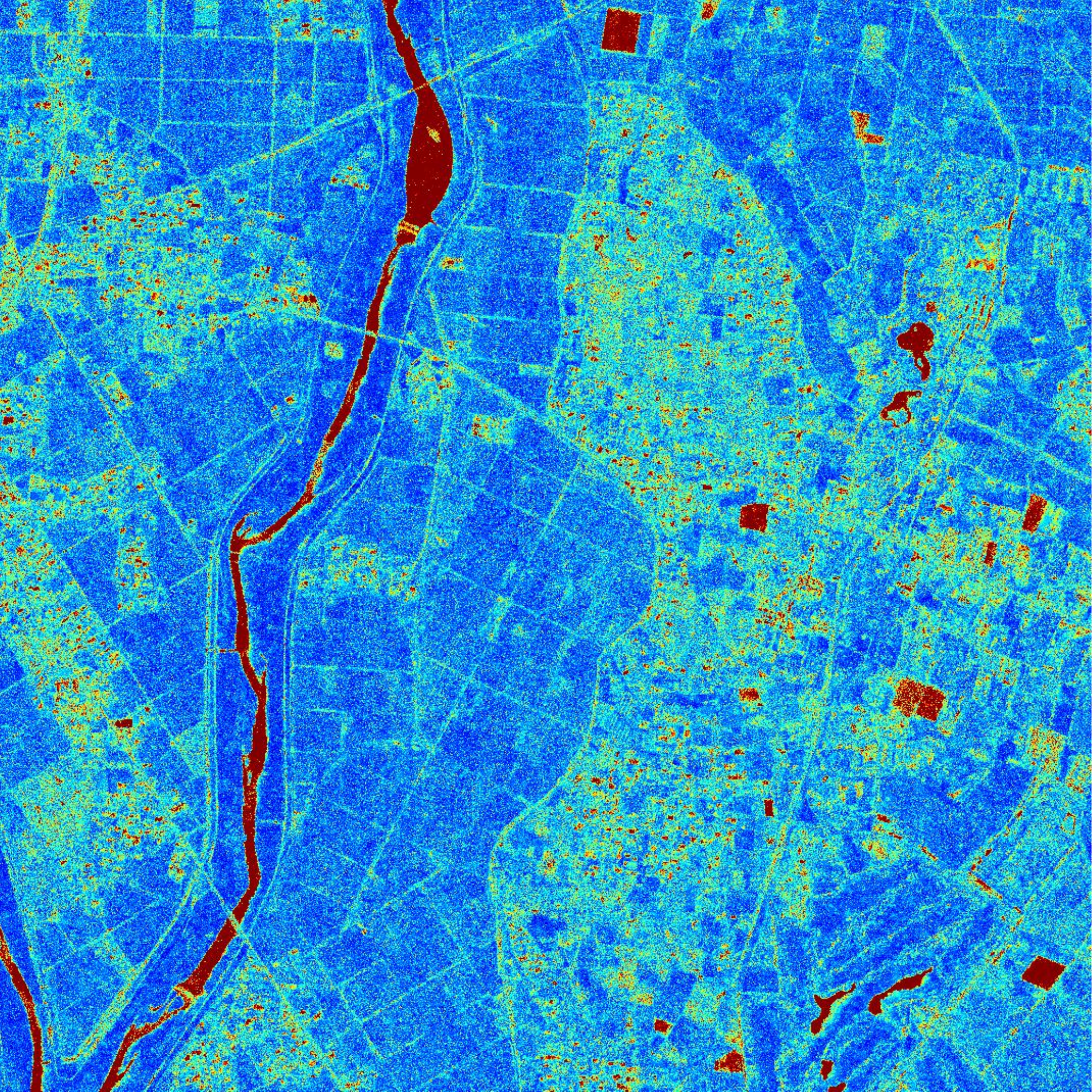}
			(b) CNMF\\
			(mean SAM $=5.59^\circ$)
		\end{minipage}
		\begin{minipage}[b]{.33\textwidth}
			\centering
			\includegraphics[width=\linewidth]{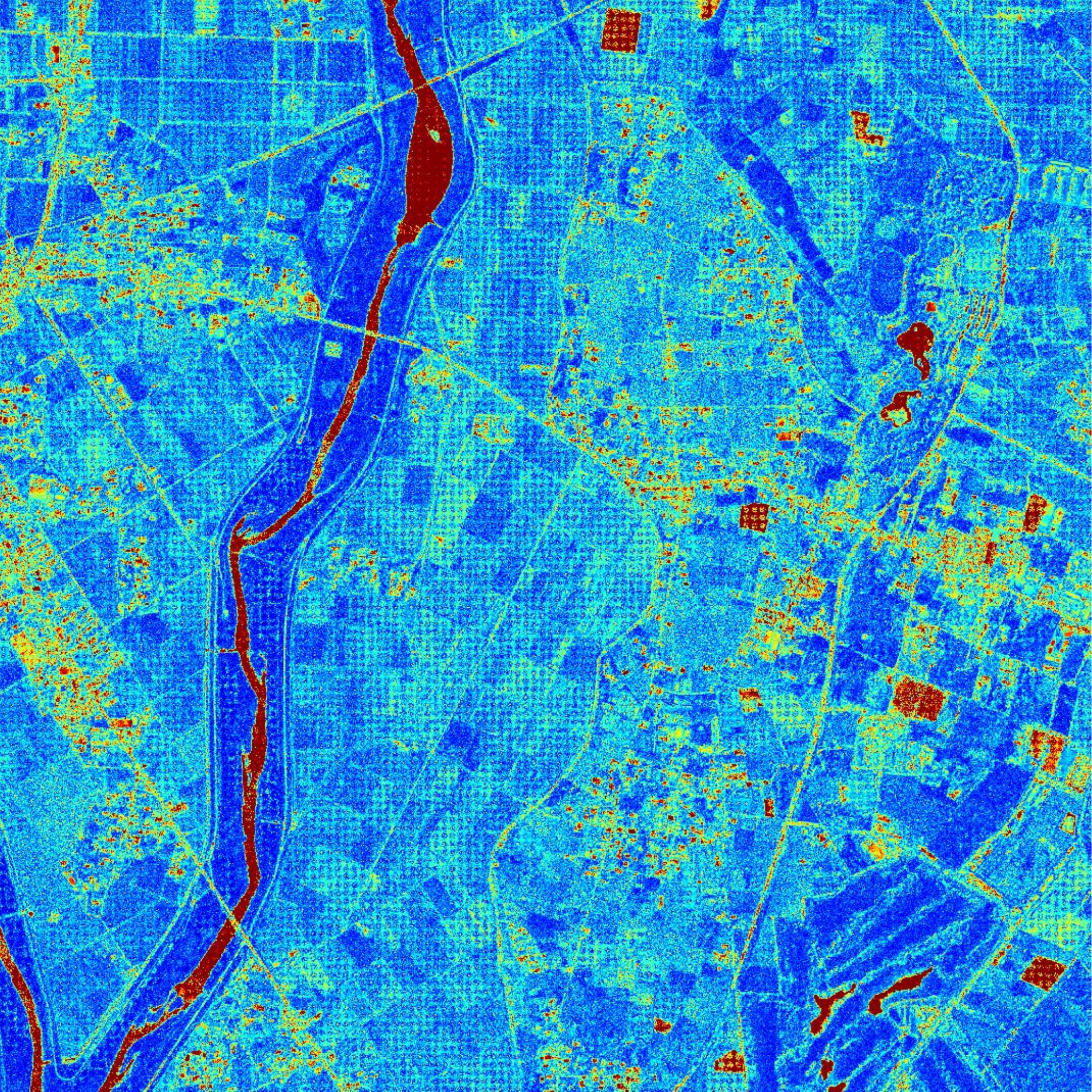}
			(c) FUMI\\
			(mean SAM $=5.47^\circ$)
		\end{minipage}
		\begin{minipage}[b]{.33\textwidth}
			\centering
			\includegraphics[width=\linewidth]{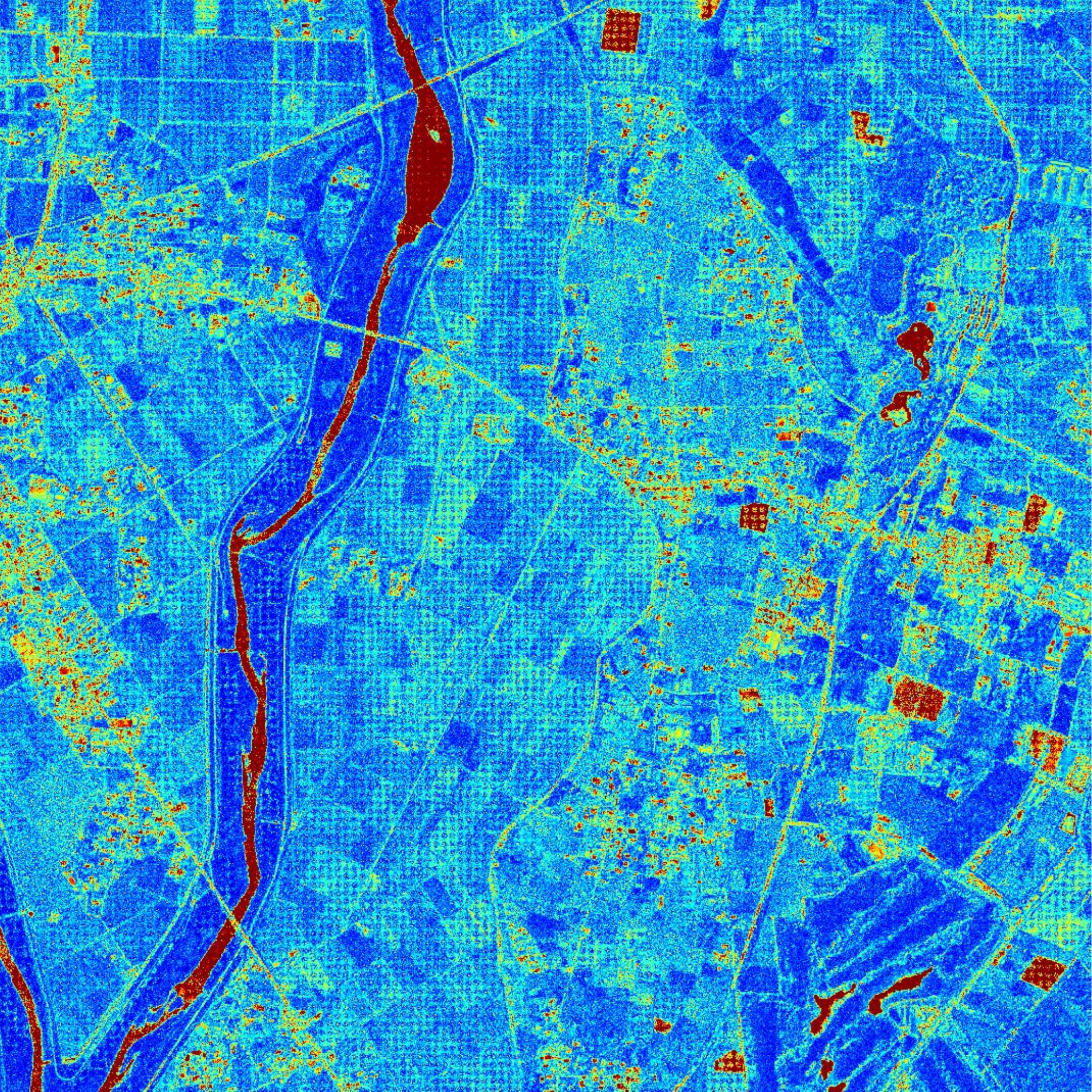}
			(d) SupResPALM \\
			(mean SAM $=4.30^\circ$)
		\end{minipage}\vspace*{.5em}
		\begin{minipage}[b]{.33\textwidth}
			\centering
			\includegraphics[width=\linewidth]{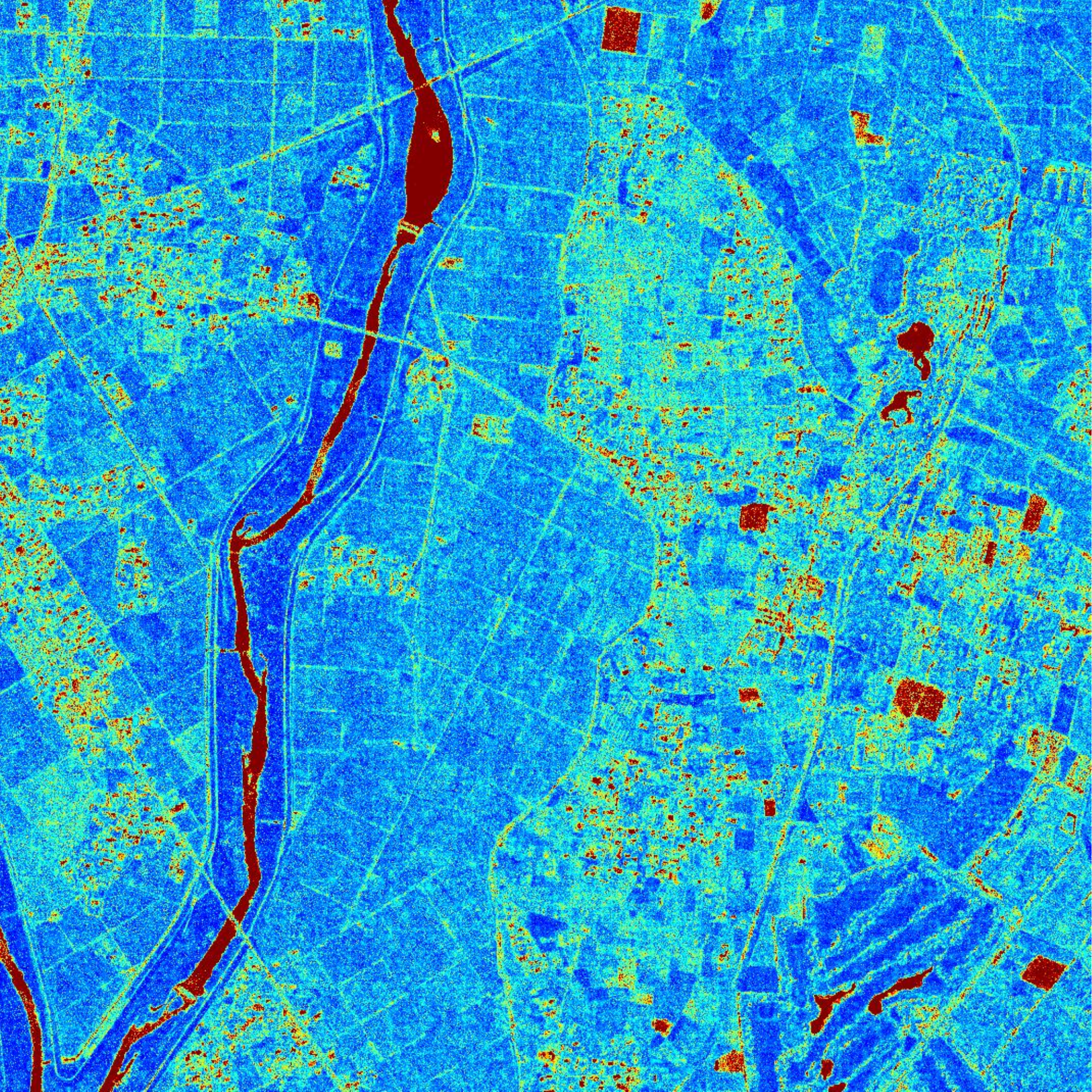}
			(e) HiBCD--plain CoSMF \\
			(mean SAM $=5.90^\circ$)
		\end{minipage}
		\begin{minipage}[b]{.33\textwidth}
			\centering
			\includegraphics[width=\linewidth]{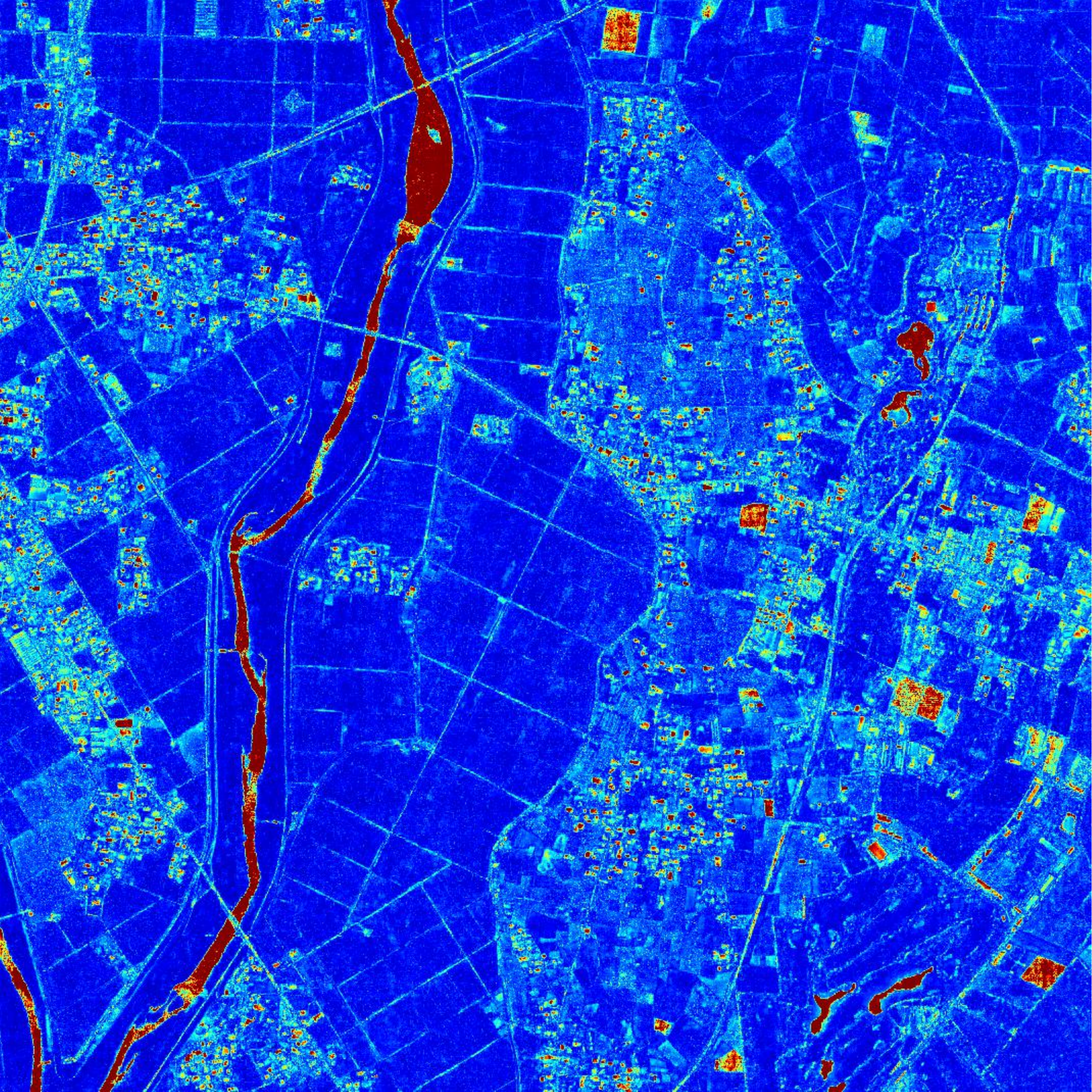}
			(f) HiBCD--NNC CoSMF \\
			(mean SAM $=3.46^\circ$)
		\end{minipage}
	\end{minipage}
	\caption{SAM maps of the various algorithms.}
	\label{fig:sam_map}
\end{figure}

\begin{figure}[h]
	\centering
	\includegraphics[width=.7\linewidth]{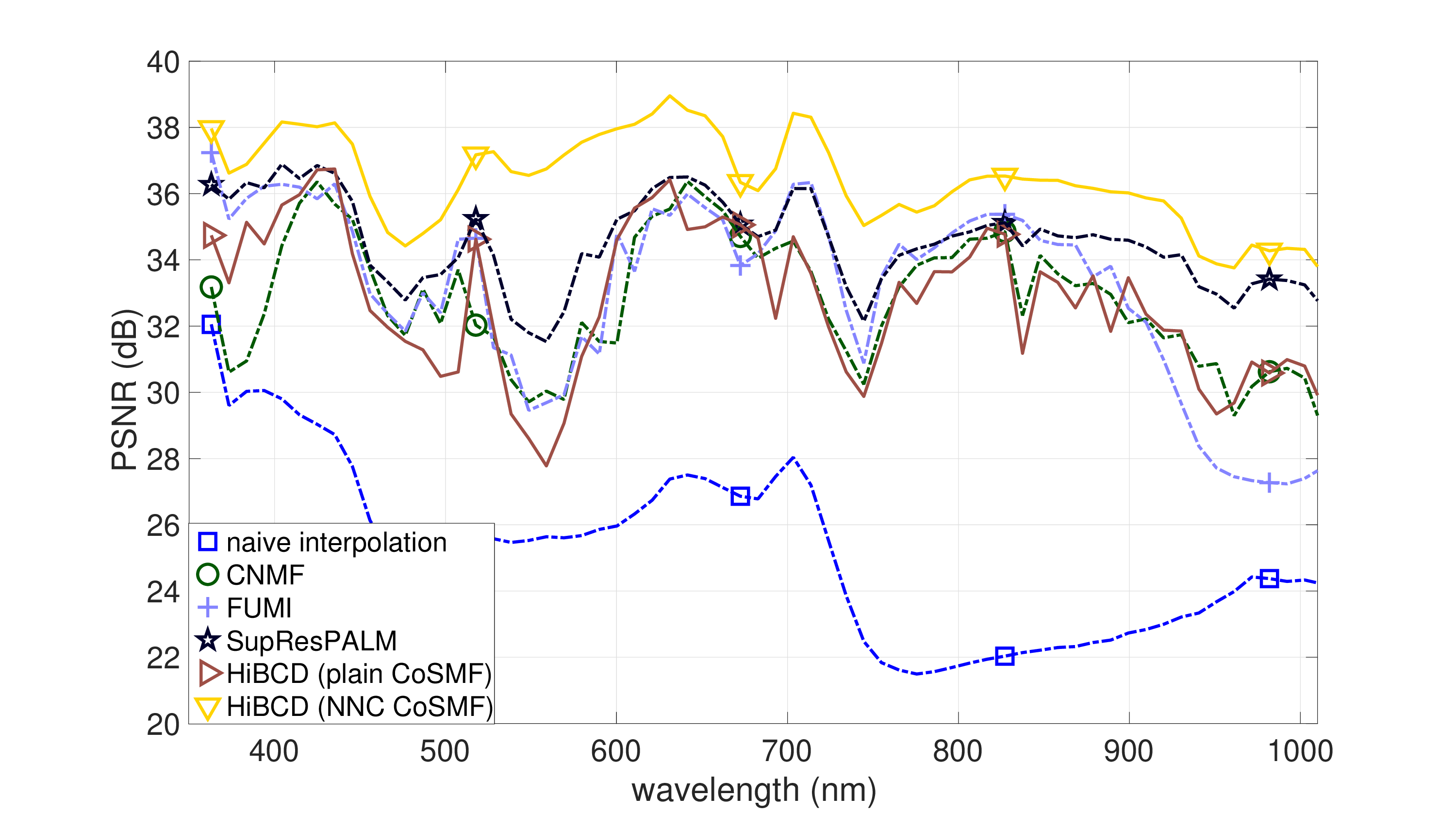}
	\caption{PSNRs versus wavelengths.}
	\label{fig:psnr_curve}
\end{figure}

We tested all the algorithms on one realization of $(\bYm,\bYh)$ (we will have Monte-Carlo  results later).
The results are shown in Figs.~\ref{fig:sam_map}--\ref{fig:psnr_curve} and Table~\ref{tab:chikusei}.
Specifically, Fig.~\ref{fig:sam_map} displays the recovery error maps of the different algorithms, evaluated by the spectral angle mapper (SAM)
\[
{\sf SAM}(\bx_i,\hat{\bx}_i) = {\rm arccos}(  \langle \bx_i, \hat{\bx}_i \rangle / (\|  \bx_i \| \| \hat{\bx}_i \|  ) ),
\]
where $\hat{\bX}$ denotes an estimated SR image;
Fig.~\ref{fig:psnr_curve} shows the peak SNRs (PSNRs) w.r.t. the spectral bands;
and Table~\ref{tab:chikusei} lists the runtimes, objective value and Erreur Relative Globale Adimensionnelle de Syth\`{e}se (ERGAS)
\[
{\sf ERGAS}(\bX,\hat{\bX}) = \frac{100}{S} \sqrt{  \frac{1}{M} \sum_{i=1}^M \frac{{\sf MSE}(\bar{\bx}_i,\hat{\bar{\bx}}_i)}{\mu^2(\hat{\bar{\bx}}_i))}  },
\]
where $S= \sqrt{\Mm/M}$ is the ratio of ground sample difference of the MS and HS images;
$\hat{\bar{\bx}}_i$ is the $i$th row of $\hat{\bX}$; $\mu(\cdot)$ and ${\sf MSE}(\cdot)$ denote the mean and mean-square error of its argument, resp.
In Table~\ref{tab:chikusei}, ``FPG-FW'' refers to the HiBCD with the FPG update for $\bA$ and the FW update for $\bS$;
``FPG-FPG'' and ``FW-FW'' are the pure FPG and FW HiBCDs, resp.

\begin{table}[h!]
	\centering
	\caption{ERGAS, runtimes and objective values of the various algorithms.}\vspace*{.5em}
	\label{tab:chikusei}
	\resizebox{.6\linewidth}{!}{%
		\begin{tabular}{|c|c|c|c|c|c|c|c|}\hline
			\multicolumn{2}{|c|}{algorithm} & ERGAS & time (sec.) & objective value  \\\hline
			\multicolumn{2}{|c|}{ideal value} & 0 & 0 & -  \\\hline
			\multicolumn{2}{|c|}{naive interpolation} & 6.96 & - & -  \\\hline
			\multicolumn{2}{|c|}{CNMF} & 3.70 & 761.12 & - \\\hline
			\multirow{4}{*}{plain} & FUMI & 3.14 & 2732.98 & 398.69 \\
			\multirow{4}{*}{CoSMF} & SupResPALM & 2.87 & 507.59 & 458.46 \\
			& FPG-FW & 3.70 & 219.02 & 375.59 \\
			& FPG-FPG & 3.50 & {\bf\red 179.80} & 367.75  \\
			& FW-FW & 3.48 & 226.55 & 378.80 \\\hline
			\multirow{2}{*}{NNC} & FPG-FW & 2.41 & {\red\bf855.55} & 791.72\\
			\multirow{2}{*}{CoSMF} & FPG-FPG & 3.42 & 3997.01 & 664.34 \\
			& FW-FW & 2.39 & 1003.54 & 809.99 \\\hline
		\end{tabular}%
	}
\end{table}

Let us discuss the results.
First, as indicated by SAM, PSNR and ERGAS, NNC CoSMF works better than plain CoSMF.
This suggests that exploiting spatial structures is beneficial for enhancing recovery performance.
Second, for plain CoSMF, SupResPALM is seen to yield better recovery performance than FUMI and HiBCD, while  FUMI and HiBCD has similar recovery performance.
Curiously, the objective value of SupResPALM is actually higher than those of FUMI and HiBCD.
While understanding the behaviors of SupResPALM is beyond the scope of this paper, we suspect that the reason is not with optimization.
Third, for plain CoSMF, the runtime performance of the three HiBCD algorithms is much better than that of FUMI and SupResPALM;
FPG-FPG is the fastest, while FPG-FW comes next.
For NNC CoSMF, FPG-FW runs faster than FPG-FPG.

To better understand the complexity of the HiBCD scheme,
we show the number of iterations and the average runtime per iteration in Table \ref{tab:time_iter_Chikusei}.
For both the plain and NNC CoSMF cases, we observe that i) FPG-FW and FW-FW have lower runtime per iteration than FPG-FPG (an expected phenomenon by design); and that ii) FPG-FPG has  smaller numbers of iterations, or converges faster, than FPG-FW and FW-FW.
For plain CoSMF, FPG-FPG has the fastest overall runtime because its advantage of smaller number of iterations compensates its weakness of longer runtime per iteration.
We however see the converse for NNC CoSMF, wherein the runtime per iteration of FPG-FPG is considerably longer than that of FPG-FW or FW-FW, and the fast convergence of FPG-FPG cannot compensate the per-iteration complexity drawback in the overall runtime.

\begin{table}[h!]
	\centering
	\caption{Complexity behaviors of the HiBCD algorithms.}\vspace*{.5em}
	\label{tab:time_iter_Chikusei}
	\resizebox{.5\linewidth}{!}{%
		\begin{tabular}{|c|c|c|c|}
			\hline
			{\bf plain CoSMF} & FPG-FW & FPG-FPG & FW-FW \\\hline
			time (sec.) & 219.02 & {\red\bf 179.80} & 226.55 \\
			iteration & 518 & {\red\bf 251} & 538 \\
			time/iteration & 0.423 & 0.716 & {\red\bf 0.421} \\\hline\hline
			{\bf NNC CoSMF} & FPG-FW & FPG-FPG & FW-FW \\\hline
			time (sec.) & {\bf\red 855.55} & 3997.01 & 1003.54 \\
			iteration & 943 & {\bf\red787} & 1109 \\
			time/iteration & 0.907 & 5.079 & {\bf\red 0.905} \\\hline
		\end{tabular}%
	}
\end{table}

A key attribute to the computational efficiency of the HiBCD scheme shown above is our custom design in Section~\ref{sec:cosmf}, especially, the step-size rule.
To give the reader some idea, we change the step-size rules to the standard ones in the literature and see what happens.
Specifically, we consider plain CoSMF and change the step-size rule for the update of $\bS$ to the standard ones in FPG and FW methods.
Table~\ref{tab:time_iter_stepsize} shows the results.
We see that our proposed methods lead to faster convergence; this is particularly so for FPG-FW.

\begin{table}[h!]
	\centering
	\caption{Impact of step-size selection for plain CoSMF.}\vspace*{.5em}
	\label{tab:time_iter_stepsize}
	\resizebox{.85\linewidth}{!}{%
		\begin{tabular}{|c|c|c|c|}
			\hline
			algorithm & step-size rule for the update of $\bS$ & time (sec.) & iteration  \\\hline
			\multirow{2}{*}{FPG-FW} &
			proposed rule \eqref{eq:FW_S_stepsize_abs} & 203.88 & {\bf\red 495} \\
			& standard rule \eqref{eq:FW_stepsize_org}  w/ $\beta=$ tight L. const. on $\Rbb^{N\times L}$ & 710.37 & 1936  \\\hline
			\multirow{2}{*}{FPG-FPG} & proposed rule \eqref{eq:beta_S_1}
			& 179.90 & {\bf\red 239}  \\
			&
			tight L. const. on $\Rbb^{N\times L}$
			& 206.98 & 276  \\\hline
		\end{tabular}%
	}
\end{table}

\subsection{Synthetic Data Experiment}

Next, we provide a Monte-Carlo simulation result using synthetic data.
The true $\bX$ is generated, at each trial, by the linear spectral mixture model \eqref{eq:lmm}.
The columns of $\bA$ are randomly chosen from a material spectral signature library, namely, the USGS library \cite{kokaly2017usgs}.
The rows of $\bS$ are randomly cropped sub-maps of the abundance maps retrieved from the  AVIRIS Cuprite dataset \cite{vane1993airborne}.
The spectral decimation matrix $\bF$ follows the Landsat MS sensor specification \cite{chander2009summary},
while the spatial decimation matrix $\bG$ corresponds to $11\times11$ Gaussian spreading (with variance $1.7^2$) and downsampling by a factor of $4$.
We have $(M, \Mm, L, \Lh, N) = (224, 6, 120^2, 30^2, 10)$.
For NNC CoSMF, we choose $\tau_i = 10$ for all $i$.

Table~\ref{tab:synthetic} shows the average recovery and runtime performance of the various algorithms over $100$ trials.
The results are generally consistent with those in the preceding semi-real experiment;
e.g., the advantage of the HiBCD scheme lies in runtime.
Additionally it is noted that when the SNR is $40$dB, plain CoSMF yields recovery performance comparable to NNC CoSMF.
Table~\ref{tab:syn_plain} shows the average number of iterations and the average runtime per iteration of the HiBCD algorithms.
We see similar complexity results as the previous (cf., Table \ref{tab:time_iter_Chikusei}).
For NNC CoSMF, we further observe that FPG-FW yields less number of iterations than FPG-FPG when the SNR is less than or equal to $30$dB.

\begin{table*}[h!]
	\centering
	\caption{Average performance of the algorithms on synthetic data.}\vspace*{.5em}
	\label{tab:synthetic}
	\resizebox{\textwidth}{!}{%
		\begin{tabular}{|c|c|c|c|c|c|c|c|c|c|}
			\hline
			\multicolumn{2}{|c|}{\multirow{2}{*}{algorithm}} & \multicolumn{4}{c}{10dB} & \multicolumn{4}{|c|}{20dB} \\\cline{3-10}
			\multicolumn{2}{|c|}{} & time (sec.) & PSNR (dB) & SAM & ERGAS & time (sec.) & PSNR (dB) & SAM & ERGAS \\\hline
			\multicolumn{2}{|c|}{ideal value} & 0 & $\infty$ & 0 & 0 & 0 & $\infty$ & 0 & 0 \\\hline
			\multicolumn{2}{|c|}{naive interpolation} & - & 13.74$\pm$0.56 & 14.15$\pm$0.07 & 6.97$\pm$0.41 & - & 22.92$\pm$0.44 & 4.74$\pm$0.10 & 2.41$\pm$0.17 \\\hline
			\multicolumn{2}{|c|}{CNMF} & 5.89$\pm$5.22 & 13.01$\pm$0.66 & 13.95$\pm$0.73 & 7.57$\pm$0.42 & 7.08$\pm$2.67 & 22.99$\pm$0.65 & 4.30$\pm$0.34 & 2.46$\pm$0.21 \\\hline
			\multirow{5}{*}{plain CoSMF} & FUMI & 4.27$\pm$1.40 & 13.02$\pm$0.65 & 14.43$\pm$0.90 & 7.87$\pm$0.40 & 11.17$\pm$2.91 & 21.85$\pm$1.18 & 5.26$\pm$0.90 & 2.88$\pm$0.42 \\
			& SupResPALM & 3.40$\pm$0.54 & 16.75$\pm$0.63 & 9.10$\pm$0.56 & 5.07$\pm$0.35 & 11.99$\pm$2.05 & 25.85$\pm$0.71 & 2.85$\pm$0.31 & 1.77$\pm$0.17 \\
			& FPG-FW & 0.81$\pm$0.22 & 13.81$\pm$0.55 & 13.15$\pm$0.48 & 6.98$\pm$0.31 & 0.94$\pm$0.24 & 22.64$\pm$0.61 & 4.65$\pm$0.23 & 2.56$\pm$0.15 \\
			& FPG-FPG & {\red\bf 0.69$\pm$0.14} & 13.11$\pm$0.54 & 14.42$\pm$0.46 & 7.55$\pm$0.36 & {\bf\red 0.79$\pm$0.13} & 21.59$\pm$0.56 & 5.42$\pm$0.20 & 2.89$\pm$0.17 \\
			& FW-FW & 1.03$\pm$0.23 & 15.74$\pm$0.50 & 11.26$\pm$0.45 & 6.04$\pm$0.32 & 1.08$\pm$0.24 & 23.89$\pm$0.57 & 4.04$\pm$0.18 & 2.24$\pm$0.12 \\\hline
			\multirow{3}{*}{NNC CoSMF} & FPG-FW & {\red\bf 1.22$\pm$0.31} & 23.74$\pm$0.96 & 2.11$\pm$0.33 & 2.14$\pm$0.25 & {\bf\red 2.72$\pm$0.77} & 29.94$\pm$0.72 & 1.38$\pm$0.21 & 1.06$\pm$0.10 \\
			& FPG-FPG & 4.69$\pm$1.04 & 23.64$\pm$0.83 & 2.77$\pm$0.34 & 2.21$\pm$0.25 & 11.44$\pm$3.21 & 29.47$\pm$0.83 & 1.64$\pm$0.21 & 1.14$\pm$0.12 \\
			& FW-FW & 1.62$\pm$0.63 & 24.50$\pm$1.25 & 2.05$\pm$0.30 & 1.98$\pm$0.31 & 3.68$\pm$1.06 & 30.26$\pm$0.71 & 1.36$\pm$0.21 & 1.03$\pm$0.10 \\\hline\hline
			\multicolumn{2}{|c|}{\multirow{2}{*}{algorithm}} & \multicolumn{4}{c}{30dB} & \multicolumn{4}{|c|}{40dB} \\\cline{3-10}
			\multicolumn{2}{|c|}{} & time (sec.) & PSNR (dB) & SAM & ERGAS & time (sec.) & PSNR (dB) & SAM & ERGAS \\\hline
			\multicolumn{2}{|c|}{ideal value} & 0 & $\infty$ & 0 & 0 & 0 & $\infty$ & 0 & 0 \\\hline
			\multicolumn{2}{|c|}{naive interpolation} & - & 28.83$\pm$0.97 & 1.91$\pm$0.20 & 1.24$\pm$0.19 & - & 30.51$\pm$1.49 & 1.24$\pm$0.26 & 1.05$\pm$0.22 \\\hline
			\multicolumn{2}{|c|}{CNMF} & 49.40$\pm$14.33 & 31.67$\pm$0.84 & 1.64$\pm$0.21 & 0.94$\pm$0.12 & 116.39$\pm$44.45 & 39.86$\pm$0.99 & 0.68$\pm$0.11 & 0.39$\pm$0.06 \\\hline
			\multirow{5}{*}{plain CoSMF} & FUMI & 7.81$\pm$1.58 & 32.67$\pm$0.79 & 1.28$\pm$0.16 & 0.81$\pm$0.09 & 6.76$\pm$1.38 & 37.93$\pm$0.75 & 0.67$\pm$0.09 & 0.47$\pm$0.06 \\
			& SupResPALM & 22.73$\pm$4.85 & 34.77$\pm$0.85 & 1.01$\pm$0.14 & 0.63$\pm$0.07 & 55.98$\pm$12.58 & 40.87$\pm$0.98 & 0.56$\pm$0.09 & 0.33$\pm$0.05 \\
			& FPG-FW & 1.60$\pm$0.48 & 31.47$\pm$0.73 & 1.68$\pm$0.15 & 0.94$\pm$0.08 & 3.82$\pm$1.22 & 39.83$\pm$1.12 & 0.69$\pm$0.11 & 0.38$\pm$0.06 \\
			& FPG-FPG & {\red\bf 1.31$\pm$0.26} & 30.14$\pm$0.78 & 2.05$\pm$0.19 & 1.11$\pm$0.12 & {\red\bf 2.46$\pm$0.51} & 38.01$\pm$1.16 & 0.91$\pm$0.17 & 0.50$\pm$0.10 \\
			& FW-FW & 1.90$\pm$0.51 & 32.21$\pm$0.69 & 1.51$\pm$0.12 & 0.85$\pm$0.07 & 4.38$\pm$1.23 & 39.96$\pm$0.96 & 0.66$\pm$0.09 & 0.37$\pm$0.05 \\\hline
			\multirow{3}{*}{NNC CoSMF} & FPG-FW & {\red\bf 5.77$\pm$1.58} & 35.19$\pm$1.14 & 0.98$\pm$0.18 & 0.61$\pm$0.10 & {\red\bf 13.58$\pm$3.84} & 38.29$\pm$1.98 & 0.78$\pm$0.18 & 0.45$\pm$0.12 \\
			& FPG-FPG & 23.18$\pm$7.04 & 35.50$\pm$0.84 & 0.95$\pm$0.13 & 0.58$\pm$0.07 & 39.18$\pm$12.19 & 40.02$\pm$1.72 & 0.67$\pm$0.15 & 0.38$\pm$0.09 \\
			& FW-FW & 8.36$\pm$2.38 & 35.10$\pm$1.11 & 0.99$\pm$0.17 & 0.62$\pm$0.10 & 19.76$\pm$5.56 & 38.29$\pm$1.85 & 0.77$\pm$0.17 & 0.45$\pm$0.11 \\\hline
		\end{tabular}%
	}
\end{table*}
\begin{table}[!th]
	\centering
	\caption{Average complexity behaviors of the various HiBCD algorithms on synthetic data.}\vspace*{1em}
	\label{tab:syn_plain}
	\resizebox{.65\linewidth}{!}{%
		\begin{tabular}{|c|c|c|c|c|}
			\hline
			\multicolumn{2}{|c|}{\bf plain CoSMF} & FPG-FW & FPG-FPG & FW-FW \\\hline
			\multirow{2}{*}{10dB} & time (sec.) & 0.81$\pm$0.22 & {\red\bf 0.69$\pm$0.14} & 1.03$\pm$0.23 \\
			& iteration & 114.66$\pm$13.92 & {\red\bf 93.45$\pm$7.67} & 144.71$\pm$8.60 \\\hline
			\multirow{2}{*}{20dB} & time (sec.)  & 0.94$\pm$0.24 & {\red\bf 0.79$\pm$0.13} & 1.08$\pm$0.24 \\
			& iteration & 134.09$\pm$11.82 & {\red\bf 107.26$\pm$7.16} & 153.02$\pm$11.78 \\\hline
			\multirow{2}{*}{30dB} & time (sec.) & 1.60$\pm$0.48 & {\red\bf 1.31$\pm$0.26} & 1.90$\pm$0.51 \\
			& iteration & 231.39$\pm$38.29 & {\red\bf179.94$\pm$16.91} & 268.25$\pm$33.21 \\\hline
			\multirow{2}{*}{40dB} & time (sec.) & 3.82$\pm$1.22 & {\red\bf2.46$\pm$0.51} & 4.38$\pm$1.23 \\
			& iteration & 551.57$\pm$123.88 & {\red\bf342.01$\pm$40.35} & 626.09$\pm$109.11 \\\hline
			\multicolumn{2}{|c|}{time/iteration} & 0.00695 & 0.00727 & 0.00704 \\\hline
			\hline
			\multicolumn{2}{|c|}{\bf NNC CoSMF} & FPG-FW & FPG-FPG & FW-FW \\\hline
			\multirow{2}{*}{10dB} & time (sec.) & {\red\bf1.22$\pm$0.31} & 4.69$\pm$1.04 & 1.62$\pm$0.63 \\
			& iteration & {\red\bf 140.44$\pm$27.20} & 144.39$\pm$27.48 & 181.23$\pm$53.45 \\\hline
			\multirow{2}{*}{20dB} & time (sec.) & {\red\bf2.72$\pm$0.77} & 11.44$\pm$3.21 & 3.68$\pm$1.06 \\
			& iteration & {\red\bf 307.05$\pm$55.53} & 357.26$\pm$96.27 & 418.48$\pm$84.23 \\\hline
			\multirow{2}{*}{30dB} & time (sec.) & {\red\bf 5.77$\pm$1.58} & 23.18$\pm$7.04 & 8.36$\pm$2.38 \\
			& iteration & {\red\bf 655.98$\pm$91.83} & 724.26$\pm$200.45 & 931.11$\pm$159.19 \\\hline
			\multirow{2}{*}{40dB} & time (sec.) & {\red\bf13.58$\pm$3.84} & 39.18$\pm$12.19 & 19.76$\pm$5.56 \\
			& iteration & 1546.65$\pm$265.72 & {\red\bf1233.88$\pm$351.53} & 2214.50$\pm$329.20 \\\hline
			\multicolumn{2}{|c|}{time/iteration} & 0.00879 & 0.03191 & 0.00892 \\\hline
		\end{tabular}%
	}
\end{table}

\section{Conclusion}
\label{sec:conclude}

In this paper we developed an efficient optimization scheme for CoSMF in HSR using hybrid FPG/FW inexact BCD.
We proved that, as an optimization framework, the limit points of the proposed scheme  are stationary.
Numerical experiments showed that the proposed scheme is computationally much more efficient than the state-of-the-art CoSMF algorithms.
The present work demonstrated the benefits of the proposed scheme under the basic CoSMF and the nuclear norm-constrained CoSMF,
and as future work it would be interesting to further explore its applications to other formulations.

\section*{Acknowledgment}

The authors thank Mr. Chun-Hei Chan and Dr. Xiao Fu who contributed to the conference version of this work.

\ifconfver
\appendix
\else
\appendix
\section*{Appendix}
\renewcommand{\thesubsection}{\Alph{subsection}}
\fi

\subsection{Proof of Fact~\ref{fac:quad_qub}} \label{app:fac:quad_qub}
The first result of Fact~\ref{fac:quad_qub} is equivalent to the statement that \eqref{eq:qub} holds for any $\bx, \by \in \setX$ if and only if $\beta \geq \lammax(\bPhi^\top \bR \bPhi)$.
Let us show this statement.
It can be verified that $\nabla f(\bx)  = \bq + \bR \bx$, and
\begin{align} \label{eq:fac:quad_qub:1}
f(\by) - f(\bx) & = \langle \nabla f(\bx), \by - \bx \rangle + \tfrac{1}{2} {(\by -\bx )^\top} \bR (\by - \bx).
\end{align}
Let $\bm \xi, \bm \gamma \in \Rbb^r$ be such that $\bx = \bPhi \bm \xi + \bd, \by = \bPhi \bm \gamma + \bd$. We have
\begin{equation} \label{eq:fac:quad_qub:2}
\begin{aligned}
 (\by -\bx )^\top \bR (\by - \bx)
&
 = ( \bm \gamma - \bm \xi  )^\top \bPhi^\top \bR \bPhi ( \bm \gamma - \bm \xi  )
\leq \lammax(\bPhi^\top \bR \bPhi) \| \bm \gamma - \bm \xi \|^2 \\
& = \lammax(\bPhi^\top \bR \bPhi) \| \bPhi( \bm \gamma - \bm \xi ) \|^2 =  \lammax(\bPhi^\top \bR \bPhi) \| \by - \bx \|^2.
\end{aligned}
\end{equation}
Note that in the third line of \eqref{eq:fac:quad_qub:2}, we have used the semi-orthogonality of $\bPhi$.
From \eqref{eq:fac:quad_qub:1}--\eqref{eq:fac:quad_qub:2} we see that \eqref{eq:qub} holds for any $\bx, \by \in \setX$ if $\beta \geq \lammax(\bPhi^\top \bR \bPhi)$.
Conversely, suppose that there exists a $\beta < \lammax(\bPhi^\top \bR \bPhi)$ such that \eqref{eq:qub} holds for some $\bx, \by \in \setX$.
Let $\bx$ be a point that lies in the relative interior of $\setX$.
By definition, there exists a radius $r > 0$ such that ${\cal B}(\bx,r) \cap {\rm aff} \, \setX \subseteq \setX$;
here, ${\cal B}(\bx,r)$ denotes the $2$-norm ball with center $\bx$ and radius $r$.
Let
\[
\by = \bx  + r \bPhi \bv,
\]
where $\bv$ is the principal eigenvector of $\bPhi^\top \bR \bPhi$, with unit $2$-norm.
It can be easily verified that $\by \in  {\cal B}(\bx,r)$ and $\by \in  {\rm aff} \, \setX$. Thus, $\by$ lies in $\setX$.
The vectors $\bx$ and $\by$ yield
\begin{equation} \label{eq:fac:quad_qub:3}
\begin{aligned}
& {(\by -\bx )^\top} \bR (\by - \bx) =  r ^2 \bv^\top \bPhi^\top \bR \bPhi \bv
 = r^2 \lammax(\bPhi^\top \bR \bPhi)
= \lammax(\bPhi^\top \bR \bPhi) \| \by - \bx \|^2.
\end{aligned}
\end{equation}
Putting \eqref{eq:fac:quad_qub:3} into \eqref{eq:fac:quad_qub:1}, we see that \eqref{eq:qub} is violated if $\beta < \lammax(\bPhi^\top \bR \bPhi)$.
Hence, we have shown that \eqref{eq:qub} holds for any $\bx, \by \in \setX$ if and only if $\beta \geq \lammax(\bPhi^\top \bR \bPhi)$.

The second result of Fact~\ref{fac:quad_qub} is straightforward. Again, let $\bm \xi, \bm \gamma \in \Rbb^r$ be such that $\bx = \bPhi \bm \xi + \bd, \by = \bPhi \bm \gamma + \bd$.
From $\nabla f(\bx)  = \bq + \bR \bx$, we readily see that
\begin{align*}
\| \nabla f(\by) - \nabla f(\bx) \| &
= \| \bR \bPhi ( \bm \gamma - \bm \xi ) \|
\leq \| \bR\bPhi \|_2 \| \bm \gamma - \bm \xi \| = \| \bR\bPhi \|_2 \| \by - \bx \|.
\end{align*}
Also, using the same proof method as above, it can be shown that equality in the above inequality is attained for some $\bx, \by \in \setX$.
Thus, $\| \bR\bPhi \|_2$ is the tight Lipschitz constant of $\nabla f$ on $\setX$.
Also, by letting $\bm \Psi \in \Rbb^{n \times (n-r)}$ be such that $\bU = [~ \bPhi ~ \bm \Psi ~]$ is orthogonal, we obtain
\begin{align*}
\| \bR\bPhi \|_2 & = \lammax^{1/2}( \bPhi^\top \bR^\top \bR\bPhi ) = \lammax^{1/2}( \bPhi^\top \bR^\top \bU \bU^\top \bR\bPhi ) \\
& = \lammax^{1/2}( \bPhi^\top \bR^\top \bPhi  \bPhi^\top \bR\bPhi + \bPhi^\top \bR^\top \bm \Psi  {\bm \Psi}^\top \bR\bPhi ) \\
& \geq  \lammax^{1/2}( \bPhi^\top \bR^\top \bPhi  \bPhi^\top \bR\bPhi) = \lammax(  \bPhi^\top \bR\bPhi),
\end{align*}
where the above inequality is due to the result $\lammax(\bA + \bB) \geq \lammax(\bA)$ for any symmetric $\bA$ and symmetric PSD $\bB$.
The proof of the second result of Fact~\ref{fac:quad_qub} is complete.


\subsection{Proof of Fact~\ref{fac:C_opt}}
\label{app:fac:C_opt}
Consider the following problem
\begin{equation} \label{eq:L_prob}
\min_{L=1,2,\ldots} ~ \frac{1}{L}  + \frac{L}{a},
\end{equation}
where $a > 0$. It can be shown that the optimal solution $L^\star$ to the above problem must satisfy $L^\star \leq \lceil \sqrt{a} \rceil$.
Let us apply this result to the minimization of $C_j$ over $L_j$. The corresponding $a$ is
\[
a = \begin{cases}\frac{ \hbetamin{j} \max\{ \eta_j \beta_j D_j^2, (M_j + l_j) D_j \} }{(M_j + l_j)^2 + 2 (D\beta)^2  }, & j\in\setI_{\sf FW}\\\frac{ \hbetamin{j} \max\{ (2\eta_j+1 + \frac{\rho_j}{\beta_j}) \beta_j D_j^2, (M_j + l_j) D_j \} }{(M_j + l_j)^2 + 2 (D\beta)^2  }, & j\in\setI_{\sf FPG}\end{cases}
.
\]
Let us first consider the case of $j\in\setI_{\sf FW}$.
If $ \eta_j \beta_j D_j^2 \leq (M_j + l_j) D_j$, we have
\begin{align*}
a  & = \frac{ \hbetamin{j}  (M_j + l_j) D_j }{(M_j + l_j)^2 + 2 (D\beta)^2  } \leq \frac{ \beta  (M_j + l_j) D }{(M_j + l_j)^2 +  (D\beta)^2  } \leq \frac{1}{2},
\end{align*}
where the first inequality is due to $D_i \leq D$ and $\hbetamin{j} \leq \beta_j \leq \beta$;
the second inequality is due to $2 ab \leq a^2 + b^2$.
If $ \eta_j \beta_j D_j^2 \geq (M_j + l_j) D_j$, we have
\begin{align*}
a  & = \frac{ \hbetamin{j} \eta_j \beta_j D_j^2 }{(M_j + l_j)^2 + 2 (D\beta)^2  } \leq \frac{ \eta_j \beta^2  D^2 }{ 2  (D\beta)^2  } = \frac{\eta_j}{2}.
\end{align*}
Combining the above two inequalities, we get
\beq \label{eq:aFW}
a \leq \frac{\eta_j}{2}, \quad j \in \setI_{\sf FW}.
\eeq
Similarly, it can be verified  that
\beq \label{eq:aFPG}
a \leq \eta_j + \frac{1}{2} + \frac{\rho_j}{\beta_j}, \quad  j\in\setI_{\sf FPG}
\eeq
The desired result in \eqref{eq:optLj} is thus obtained.

Moreover, we observe from Problem~\eqref{eq:L_prob} that if
\beq
1 + \frac{1}{a} \leq \frac{1}{2} + \frac{2}{a} \quad \Longleftrightarrow \quad a \leq 2,\notag
\eeq
then $L_j^\star = 1$.
It follows from \eqref{eq:aFW} and \eqref{eq:aFPG} that $L_j^\star = 1$ is true if $\frac{\eta_j}{2} \leq 2$ for $j \in \setI_{\sf FW}$, or if  $\eta_j + \frac{1}{2} \leq 2$ for $j\in\setI_{\sf FPG}$, $\rho_j= 0$.
The proof is complete.


\subsection{Proof of Lemma~\ref{lem:EPG_suff_dec}} \label{app:fpg}
We first prove \eqref{eq:EPG_suff_dec1}. From \eqref{eq:fpg_step_full}, we obtain
\beq \label{eq:lips_fpg}
\begin{split}
    f( \bx^+ ) & \leq f( \bz ) + \langle \grd_i f(\bz), \bx_i^+ - \bz_i \rangle + \frac{\hat{\beta}_i}{2} \| \bx_i^+ - \bz_i \|^2
\end{split}
\eeq
Combining \eqref{eq:lips_fpg} with the definition of the proximal operator, it can be shown that for all $\by_i \in \setX_i$,
\begin{gather}
\textstyle F( \bx^+ ) \leq f( \bz ) + \varphi_i(\by_i; \bz, \hbetai) + \sum_{j \neq i} h_j( \bx_j),  \label{eq:inter_fpg1}  \\
\varphi_i(\by_i; \bz, \hbetai) \eqdef \langle \nabla_i f(\bz), \by_i - \bz_i \rangle + \frac{\hbetai}{2} \| \by_i  - \bz_i \|^2 + h_i(\by_i). \label{eq:varphi}
\end{gather}
By noting that $f(\bx) + \frac{\rho_i}{2} \| \bx_i \|^2$ is convex in $\bx_i \in \tilde{\setX}_i$, and by applying the first-order condition for convex functions, it can be shown that
\begin{equation} \label{eq:weakcvx}
f(\bx) \geq f(\bz) + \langle \nabla_i f(\bz), \bx_i - \bz_i \rangle - \frac{\rho_i }{2} \| \bx_i - \bz_i \|^2.
\end{equation}
Upon substituting \eqref{eq:weakcvx} into \eqref{eq:inter_fpg1}, we obtain
\beq \notag
\begin{split}
    F(\bx^+) & \leq f( \bx ) + h_i ( \by_i ) + \sum_{j \neq i } h_j ( \bx_j ) + \frac{\hat{\beta}_i}{2} \| \by_i - \bz_i \|^2
    \\
    &
    + \langle \grd_i f( \bz ) , \by_i - \bx_i \rangle + \frac{\rho_i }{2} \| \bx_i - \bz_i \|^2.
\end{split}
\eeq
Let us deal with the terms in the above one by one.
We have
\begin{align*}
\langle \nabla_i f(\bz) - \nabla_i f(\bx), \by_i - \bx_i \rangle
& \leq \frac{\beta_i}{2} (\| \bz_i - \bx_i \|^2 +  \| \by_i - \bx_i \|^2 ),
\end{align*}
which is
due to the Lipschitz continuity of $\nabla_i f$, $\bz_{-i} = \bx_{-i}$, the Cauchy-Schwartz inequality and  Young's inequality.
Also,
\begin{align*}
\| \by_i  - \bz_i \|^2 &
\leq 2 \| \by_i - \bx_i \|^2_2 + 2 \| \bx_i - \bz_i \|^2,
\end{align*}
which is due to $\| \ba + \bb \|^2 \leq 2 \| \ba \|^2 + 2 \| \bb \|^2$. Letting $\bar{\beta}_i = 2 \hat{\beta}_i + \beta_i + \rho_i$, one can verify that
\begin{equation} \notag
F(\bx^+) \leq f(\bx) + \varphi_i(\by_i; \bx, \bar{\beta}_i )  +  \frac{\bar{\beta}_i}{2} \| \bx_i  - \bz_i \|^2 + \sum_{j \neq i} h_j( \bx_j),
\end{equation}
for any $\by_i \in \setX_i$.
Observe the following lemma:
\begin{Lemma} \label{lem:varphi}
    There exists a $\by_i \in \setX_i$ such that
    \begin{equation} \label{eq:lem1_meq22}
    \varphi_i(\by_i;\bx,\bar{\beta}_i) - h_i(\bx_i) \leq - \frac{g_i(\bx)^2}{2   \max\{  \bar{\beta}_i D_i^2, (M_i + l_i)D_i \} },
    \end{equation}
    where $\varphi_i(\by_i;\bx,\bar{\beta}_i)$ was defined in \eqref{eq:varphi}.
\end{Lemma}
The proof is shown in
Appendix~\ref{app:aux}.
Applying Lemma~\ref{lem:varphi} and together with the fact that $\bz_i - \bx_i = \alpha_i ( \bx_i - \bx_i^- )$, we arrive at the desired result in \eqref{eq:EPG_suff_dec1}.


To prove the second result in \eqref{eq:EPG_suff_dec2}, we note that a similar result has been shown in \cite[Lemma 2.2]{xu2013block} for the convex case. Here we extend it to the weakly-convex case. We begin by 
\begin{align}
 F(\bx) - F(\bx^+)
 \overset{(a)}{\geq} & f( \bx ) - f( \bz ) - \langle \nabla_i f( \bz ) , \bx_i - \bx_i +
\bx_i^+ - \bz_i \rangle
- \frac{\hat\beta_i}{2} \| \bx_i^+ - \bz_i  \|^2 + h_i( \bx_i ) - h_i( \bx_i^+ ) \nonumber \\
 \overset{(b)}{\geq}& - \frac{ \rho_i  }{2} \| \bx_i - \bz_i \|^2 - \frac{\hat\beta_i}{2} \| \bx_i^+ - \bz_i  \|^2
- \langle \nabla_i f( \bz ) , \bx_i^+ - \bx_i  \rangle + h_i( \bx_i ) - h_i( \bx_i^+ ), \label{eq:FPG_stuff1}
\end{align}
where (a) and (b) are due to  \eqref{eq:lips_fpg} and \eqref{eq:weakcvx}, resp.
Since $\bx_i^+$ is the output of the proximal operator, one has
\begin{equation*}
\left\langle \nabla_i f( \bz ) + \hbetai ( \bx_i^+ - \bz_i ) + {\bm g}_i, \by_i - \bx_i^+  \right\rangle \geq 0,~\forall~\by_i \in
\setX_i,
\end{equation*}
where ${\bm g}_i \in \partial h_i( \bx_i^+ )$. Setting $\by_i = \bx_i$ and substituting the above into \eqref{eq:FPG_stuff1} gives the lower bound
\begin{equation*}
\begin{split}
& F(\bx) - F(\bx^+) \geq
- \frac{ \rho_i }{2} \| \bx_i - \bz_i \|^2 - \frac{\hat\beta_i}{2} \| \bx_i^+ - \bz_i  \|^2
+ \langle \hbetai ( \bx_i^+ - \bz_i ) + {\bm g}_i, \bx_i^+ - \bx_i \rangle + h_i( \bx_i ) - h_i( \bx_i^+ ).
\end{split}
\end{equation*}
Since $h_i$ is convex, we have $h_i( \bx_i ) \geq h_i( \bx_i^+ ) + \langle {\bm g}_i, \bx_i - \bx_i^+ \rangle$, and consequently,
\begin{equation*} \label{eq:FPG_stuff2}
\begin{split}
& \langle \hbetai ( \bx_i^+ - \bz_i ) + {\bm g}_i, \bx_i^+ - \bx_i \rangle + h_i( \bx_i ) - h_i( \bx_i^+ )
\geq  \hbetai \langle  \bz_i - \bx_i^+ ,  \bx_i - \bx_i^+ \rangle. \\
\end{split}
\end{equation*}
We notice that $\langle  \bz_i - \bx_i^+ ,  \bx_i - \bx_i^+ \rangle = \frac{1}{2} \big( \| \bz_i - \bx_i^+ \|^2 + \| \bx_i - \bx_i^+ \|^2 - \| \bz_i  -\bx_i \|^2 \big)$. We obtain the  lower bound as:
\begin{equation*}
F(\bx) - F(\bx^+) \geq \frac{\hbetai}{2} \| \bx_i - \bx_i^+ \|^2 - \Big( \frac{\hbetai}{2} + \frac{\rho_i}{2} \Big)  \| \bx_i - \bz_i \|^2 .
\end{equation*}
Finally, applying $\bz_i - \bx_i = \alpha_i ( \bx_i - \bx_i^- )$ gives \eqref{eq:EPG_suff_dec2}.


\subsection{Proof of Lemma~\ref{lem:thm_step2}}\label{app:lemstep2}
Observe that
\begin{align}
 g_i(\bx^k)^2
  =& g_i(\bxk{1}{0})^2  \nonumber \\
 =& \left\{ \frac{1}{L_i} \sum_{\ell=0}^{L_i-1} [ g_i(\bxk{1}{0}) - g_i(\bxk{i}{\ell})  ]  + \frac{1}{L_i} \sum_{\ell=0}^{L_i-1} g_i(\bxk{i}{\ell})  \right\}^2 \nonumber  \\
 \leq& \frac{2}{L_i} \left\{ \sum_{\ell=0}^{L_i-1} [ g_i(\bxk{1}{0}) - g_i(\bxk{i}{\ell})  ]^2 + \sum_{\ell=0}^{L_i-1} g_i(\bxk{i}{\ell})^2 \right\} \label{eq:thm1_step2_eq1}
\end{align}
where the above inequality is due to $(\sum_{i=1}^n z_i)^2 \leq n \sum_{i=1}^n z_i^2$.
Let us deal with the terms in \eqref{eq:thm1_step2_eq1} one by one.
First, from \eqref{eq:thm1_step1_eq1} and from the definitions of $\dF{k,i}$ and $\tdx{k,i}$, we see that
\begin{align}
\sum_{\ell=0}^{L_i-1} g_i(\bxk{i}{\ell})^2 & \leq A_i  ( \dF{k,i} + \tdx{k,i}  ),
\label{eq:thm1_step2_eq2}
\end{align}
Second, we have
\begin{align}
[ g_i(\bxk{1}{0}) - g_i(\bxk{i}{\ell})  ]^2
& \leq
\left[ (D_i \beta) \| \bxk{1}{0} - \bxk{i}{\ell} \|_2 + (M_i + l_i) \| \bxk{1}{0}_i - \bxk{i}{\ell}_i \|_2  \right]^2 \nonumber  \\
& \leq 2 (D_i \beta)^2 \| \bxk{1}{0} - \bxk{i}{\ell} \|^2 + 2(M_i + l_i)^2 \| \bxk{1}{0}_i - \bxk{i}{\ell}_i \|^2, \notag
\end{align}
where the first inequality is due to Lemma~\ref{fact:Lip_g}. Moreover,
\begin{align*}
\| \bxk{1}{0} - \bxk{i}{\ell} \|^2 & = \sum_{j=1}^{i} \| \bxk{1}{0}_j - \bxk{i}{\ell}_j \|^2,
\end{align*}
\beq \notag
\| \bxk{1}{0}_j - \bxk{i}{\ell}_j \|^2 = \begin{cases}
    \| \bxk{j}{0}_j - \bxk{j}{L_j}_j \|^2,~&\text{if}~j < i,\vspace{.1cm} \\
    \| \bxk{i}{0}_i - \bxk{i}{\ell}_i \|^2,~&\text{if}~j = i.
\end{cases}
\eeq
Also, let us define $\dx{k,j} \eqdef \sum_{\ell'=0}^{L_j-1} \dx{k,j,\ell'}$. We have that
\begin{align*}
& \| \bxk{1}{0}_j - \bxk{i}{\ell}_j \|^2 = \left\|  \sum_{\ell'=0}^{L_i-1} ( \bxk{j}{\ell'}_j -\bxk{j}{\ell'+1}_j ) \right\|^2
\leq L_j {\color{black} \sum_{\ell'=0}^{L_i-1} \| \bxk{j}{\ell'}_j -\bxk{j}{\ell'+1}_j  \|^2 = L_j \dx{k,j}},
\end{align*}
and similarly one has $\| \bxk{1}{0}_i - \bxk{i}{\ell}_i \|^2 \leq \ell \dx{k,i}$.
The above results give:
\begin{align}
[ g_i(\bxk{1}{0}) - g_i(\bxk{i}{\ell})  ]^2
& \leq 2 (D_i \beta )^2 \sum_{j=1}^{i-1} L_j \dx{k,j} + 2 [ (D_i \beta )^2 + (M_i + l_i)^2 ] \ell \dx{k,i}
\label{eq:thm1_step2_eq4}
\end{align}
Third, summing \eqref{eq:thm1_step2_eq4} w.r.t. $\ell$ yields
\begin{align}
& \frac{1}{L_i} \sum_{\ell=0}^{L_i - 1} [ g_i(\bxk{1}{0}) - g_i(\bxk{i}{\ell})  ]^2
\leq \sum_{j=1}^m \tilde{C}_{ij} \dx{k,j},
\label{eq:thm1_step2_eq5}
\end{align}
where \[
\tilde{C}_{ij}  = \left\{
\begin{array}{ll}
\displaystyle 2 (D_i \beta)^2 L_j, & 1 \leq j \leq i-1 \\
\displaystyle [ (D_i \beta)^2 + (M_i + l_i)^2 ] (L_i - 1), & i=j \\
0, & \text{otherwise}
\end{array}
\right.
\] Here,
we have used $\sum_{\ell=0}^{L_i-1} \ell = (L_i-1)L_i/2$ to obtain \eqref{eq:thm1_step2_eq5}.
From \eqref{eq:thm1_step2_eq5} we further derive
\begin{align}
\sum_{j=1}^m \tilde{C}_{ij} \dx{k,j} & \leq
\sum_{j=1}^m \frac{ 2\tilde{C}_{ij} }{\hbetamin{j}}
\left( \sum_{\ell=0}^{L_j-1} \frac{\hat{\beta}_{j}^{k,\ell} }{2} \dx{k,j,\ell} \right)
\nonumber \\
& \leq \sum_{j=1}^m \frac{ 2\tilde{C}_{ij} }{\hbetamin{j}}
\left[ \sum_{\ell=0}^{L_j-1} \left( \dF{k,j,\ell} +  \frac{\bar{\beta}_{j}^{k,\ell} (\alpha_{j}^{k,\ell})^2}{2} \dx{k,j,\ell} \right) \right] \nonumber \\
& = \sum_{j=1}^m \frac{ 2\tilde{C}_{ij} }{\hbetamin{j}}  ( \dF{k,j} + \tdx{k,j} ),
\label{eq:thm1_step2_eq5_5}
\end{align}
where
the second inequality is due to \eqref{eq:thm1_step1_eq2}, $\bar{\beta}_{j}^{k,\ell} \geq \hat{\beta}_{j}^{k,\ell}$ for $j \in \setI_{\sf FPG}$,
and $\alpha_j^{k,\ell} =0$ for $j \notin \setI_{\sf FPG}$.
Finally, by putting \eqref{eq:thm1_step2_eq2}, \eqref{eq:thm1_step2_eq5} and \eqref{eq:thm1_step2_eq5_5} into \eqref{eq:thm1_step2_eq1}, we get
\begin{align*}
g_i(\bx^k) & \leq \sum_{j=1}^m \frac{ 4\tilde{C}_{ij} }{\hbetamin{j}}  ( \dF{k,j} + \tdx{k,j} ) + \frac{2 A_i}{L_i} ( \dF{k,i} + \tdx{k,i} ).
\end{align*}
By defining $C_{ii} = 4\tilde{C}_{ii}/\hbetamin{i} + 2 A_i/L_i$ and $C_{ij} =4\tilde{C}_{ij}/\hbetamin{j}$ for $i \neq j$,
we obtain the desired result in  \eqref{eq:thm_step2_main_eq}.
The proof is done.

\subsection{Proof of Eq.~\eqref{eq:fpgtelescope}} \label{app:fpgtele}
From \eqref{eq:thm1_step1_eq2}, we see that for $k \geq 1$,
\begin{align*}
\dF{k,i} & \geq \sum_{\ell=0}^{L_i-1}  \frac{\hat{\beta}_{i}^{k,\ell}}{2} \big\{ \dx{k,i,\ell} - \big(1+ \frac{\rho_i}{\hat{\beta}_i^{k,\ell}} \big) (\alpha_i^{k,\ell})^2 \dx{k,i,\ell-1} \big\} \nonumber \\
& = \sum_{\ell=0}^{L_i-1}  \Big( \frac{\hat{\beta}_{i}^{k,\ell-1}}{2} - \big(1+ \frac{\rho_i}{\hat{\beta}_i^{k,\ell}} \big) \frac{\hat{\beta}_{i}^{k,\ell} (\alpha_i^{k,\ell})^2 }{2}  \Big) \dx{k,i,\ell-1}
+ \frac{\hat{\beta}_{i}^{k,L_i-1}}{2} \dx{k,i,L_i-1} - \frac{\hat{\beta}_{i}^{k,-1}}{2} \dx{k,i,-1}
\end{align*}
and similarly for $k=0$,
\begin{align*}
\dF{k,i} & \geq
\sum_{\ell=0}^{L_i-1}  \Big( \frac{\hat{\beta}_{i}^{k,\ell-1}}{2} - \big( 1+ \frac{\rho_i}{\hat{\beta}_i^{k,\ell}}  \big) \frac{\hat{\beta}_{i}^{k,\ell} (\alpha_i^{k,\ell})^2 }{2}  \Big) \dx{k,i,\ell-1}
 + \frac{\hat{\beta}_{i}^{k,L_i-1}}{2} \dx{k,i,L_i-1}
\end{align*}
From the extrapolation weight condition \eqref{eq:alpha_cond}, we observe that
\begin{align}
{\hat{\beta}_{i}^{k,\ell-1}} - \big(1+ \frac{\rho_i}{\hat{\beta}_i^{k,\ell}}  \big) {\hat{\beta}_{i}^{k,\ell} (\alpha_i^{k,\ell})^2 } & \geq
\left( \frac{1}{ \bar{\aaalp{}}^2  } - 1 \right) \hat{\beta}_{i}^{k,\ell} (\alpha_i^{k,\ell})^2.\notag
\end{align}
Note that \eqref{eq:alpha_cond} holds for any $i \in\{1,...,m\}$ using the extended definition that $\alpha_i^{k,\ell} = 0$ if $i \in \setI_{\sf FW}$, defined after \eqref{eq:thm1_step1_eq2}.
We have
\begin{align}
\sum_{k=0}^K \sum_{i=1}^m \dF{k,i} & \geq
\sum_{k=0}^K \sum_{i=1}^m \sum_{\ell=0}^{L_i-1} \left( \frac{1}{ \bar{\aaalp{}}^2  } - 1 \right) \frac{\hat{\beta}_{i}^{k,\ell} (\alpha_i^{k,\ell})^2 }{2} \dx{k,i,\ell-1} \label{eq:thm_step4_eq1} 
\end{align}
Furthermore, we see from the definition of $\tdx{k,i}$ that
\begin{align}
\tdx{k,i} & = \sum_{\ell=0}^{L_i-1} \frac{( 2 \hat{\beta}_{i}^{k,\ell}  + \beta_i + \rho_i ) (\alpha_i^{k,\ell})^2}{2} \dx{k,i,\ell-1}
\leq \left(2 + \frac{\beta_i + \rho_i}{\hbetamini} \right) \sum_{\ell=0}^{L_i-1} \frac{ \hat{\beta}_{i}^{k,\ell} (\alpha_i^{k,\ell})^2}{2} \dx{k,i,\ell-1}
\label{eq:thm_step4_eq4}
\end{align}
Substituting \eqref{eq:thm_step4_eq4} into \eqref{eq:thm_step4_eq1} and noticing that $\alpha_i^{k,\ell} = 0$ if $i \in \setI_{\sf FW}$, we obtain the desired result in \eqref{eq:fpgtelescope}.

\subsection{Proof of Lemma~\ref{lem:varphi}}\label{app:aux}

To show the inequality, let
$$\bp_i \in \arg \max_{\by_i \in \setX_i} \langle \nabla_i f(\bx), \bx_i - \by_i \rangle + h_i(\bx_i) - h_i(\by_i).$$
Note that
\beq
g_i(\bx) = \langle \nabla_i f(\bx), \bx_i - \bp_i \rangle + h_i(\bx_i) - h_i(\bp_i) \label{eq:hello}.
\eeq
Let
\begin{equation}
\by_i = (1-t) \bx_i + t \bp_i,\notag
\end{equation}
for some $t \in [0,1]$.
We get
\[
\varphi_i(\by_i;\bx,\hat{\beta}_i) = -t \langle \nabla_i f(\bx), \bx_i - \bp_i \rangle + \frac{\hat{\beta}_i t^2}{2} \| \bp_i - \bx_i \|^2 + h_i(\by_i),
\]
By applying $h_i(\by_i) \leq (1 -t )h_i(\bx_i) + t \, h_i(\bp_i)$ (because $h_i$ is convex) to the above equation, we further obtain
\begin{equation} \label{eq:lem1_whatever}
\varphi_i(\by_i;\bx,\hat{\beta}_i) - h_i(\bx_i) \leq r(t):=  - t g_i(\bx) + \frac{\hat{\beta}_i t^2}{2} \| \bp_i - \bx_i \|^2
\end{equation}
The remaining proof is similar to part of the proof in \cite[Lemma 4.6--4.7]{beck2015cyclic}.
For self-containedness we concisely describe the proof.
Let $t=  \min\{ 1, g_i(\bx)/[ \hat{\beta}_i \| \bp_i - \bx_i \|^2 ] \}$.
If $t < 1$, we have $t= g_i(\bx)/[ \hat{\beta}_i \| \bp_i - \bx_i \|^2 ]$.
Putting this $t$ into $r(t)$  yields
\begin{equation}
\label{eq:lem1_beckold2}
r(t) = - \frac{g_i(\bx)^2}{ 2 \hat{\beta}_i \| \bp_i - \bx_i \|^2 }
\leq  - \frac{g_i(\bx)^2}{ 2 \hat{\beta}_i D_i^2 }.
\end{equation}
If $t= 1$, we have $g_i(\bx) \geq \hat{\beta}_i \| \bp_i - \bx_i \|^2$.
Using this inequality, and putting $t=1$ into $r(t)$, we get
\begin{equation} \label{eq:lem1_beckold3}
r(t) = r(1)   \leq -\tfrac{1}{2} g_i(\bx) \leq - \frac{g_i(\bx)^2}{2(M_i+l_i)D_i},
\end{equation}
where the second inequality is due to $g_i(\bx) \leq (M_i+l_i)D_i$; specifically, from \eqref{eq:hello},
\begin{align*}
g_i(\bx) & \leq \| \nabla_i f(\bx) \| \| \bx_i - \bp_i \| + l_i \| \bx_i - \bp_i \|
\leq (M_i + l_i ) D_i,
\end{align*}
in which the first inequality is due to the Cauchy-Schwartz inequality and the Lipschitz continuity of $h_i$ on $\setX_i$, and we recall $D_i = \max_{\bx_i,\by_i \in \setX_i} \| \bx_i - \by_i \|$, $M_i = \max_{ \bx \in \setX_i } \| \grd_i f(\bx) \|$.
It follows from \eqref{eq:lem1_whatever}--\eqref{eq:lem1_beckold3} that \eqref{eq:lem1_meq22} is true. 


\bibliographystyle{IEEEtran}
\bibliography{refs_hsi,refs_opt}

\begin{thebibliography}{10}
\providecommand{\url}[1]{#1}
\csname url@samestyle\endcsname
\providecommand{\newblock}{\relax}
\providecommand{\bibinfo}[2]{#2}
\providecommand{\BIBentrySTDinterwordspacing}{\spaceskip=0pt\relax}
\providecommand{\BIBentryALTinterwordstretchfactor}{4}
\providecommand{\BIBentryALTinterwordspacing}{\spaceskip=\fontdimen2\font plus
\BIBentryALTinterwordstretchfactor\fontdimen3\font minus
  \fontdimen4\font\relax}
\providecommand{\BIBforeignlanguage}[2]{{%
\expandafter\ifx\csname l@#1\endcsname\relax
\typeout{** WARNING: IEEEtran.bst: No hyphenation pattern has been}%
\typeout{** loaded for the language `#1'. Using the pattern for}%
\typeout{** the default language instead.}%
\else
\language=\csname l@#1\endcsname
\fi
#2}}
\providecommand{\BIBdecl}{\relax}
\BIBdecl

\bibitem{loncan2015hyperspectral}
L.~Loncan, L.~B. De~Almeida, J.~M. Bioucas-Dias, X.~Briottet, J.~Chanussot,
  N.~Dobigeon, S.~Fabre, W.~Liao, G.~A. Licciardi, M.~Simoes \emph{et~al.},
  ``Hyperspectral pansharpening: {A} review,'' \emph{IEEE Geosci. Remote Sens.
  Mag.}, vol.~3, no.~3, pp. 27--46, 2015.

\bibitem{yokoya2017hyperspectral}
N.~Yokoya, C.~Grohnfeldt, and J.~Chanussot, ``Hyperspectral and multispectral
  data fusion: A comparative review of the recent literature,'' \emph{IEEE
  Geosci. Remote Sens. Mag.}, vol.~5, no.~2, pp. 29--56, 2017.

\bibitem{palsson2017multispectral}
F.~Palsson, J.~R. Sveinsson, and M.~O. Ulfarsson, ``Multispectral and
  hyperspectral image fusion using a 3-{D}-convolutional neural network,''
  \emph{IEEE Geosci. Remote Sens. Lett.}, vol.~14, no.~5, pp. 639--643, 2017.

\bibitem{lanaras2018super}
C.~Lanaras, J.~Bioucas-Dias, S.~Galliani, E.~Baltsavias, and K.~Schindler,
  ``Super-resolution of {S}entinel-2 images: Learning a globally applicable
  deep neural network,'' \emph{ISPRS Journal of Photogrammetry and Remote
  Sensing}, vol. 146, pp. 305--319, 2018.

\bibitem{kawakami2011high}
R.~Kawakami, Y.~Matsushita, J.~Wright, M.~Ben-Ezra, Y.-W. Tai, and K.~Ikeuchi,
  ``High-resolution hyperspectral imaging via matrix factorization,'' in
  \emph{Proc. IEEE Conf. Comput. Vis. Pattern Recognit.}, 2011, pp. 2329--2336.

\bibitem{yokoya2012coupled}
N.~Yokoya, T.~Yairi, and A.~Iwasaki, ``Coupled nonnegative matrix factorization
  unmixing for hyperspectral and multispectral data fusion,'' \emph{IEEE Trans.
  Geosci. Remote Sens.}, vol.~50, no.~2, pp. 528--537, 2012.

\bibitem{lanaras2015hyperspectral}
C.~Lanaras, E.~Baltsavias, and K.~Schindler, ``Hyperspectral super-resolution
  by coupled spectral unmixing,'' in \emph{Proc. IEEE Int. Conf. Computer
  Vision}, 2015, pp. 3586--3594.

\bibitem{wei2016multiband}
Q.~Wei, J.~Bioucas-Dias, N.~Dobigeon, J.-Y. Tourneret, M.~Chen, and S.~Godsill,
  ``Multiband image fusion based on spectral unmixing,'' \emph{IEEE Trans.
  Geosci. Remote Sens.}, vol.~54, no.~12, pp. 7236--7249, 2016.

\bibitem{simoes2015convex}
M.~Sim{\~o}es, J.~Bioucas-Dias, L.~B. Almeida, and J.~Chanussot, ``A convex
  formulation for hyperspectral image superresolution via subspace-based
  regularization,'' \emph{IEEE Trans. Geosci. Remote Sens.}, vol.~53, no.~6,
  pp. 3373--3388, 2015.

\bibitem{lanaras2017hyperspectral}
C.~Lanaras, E.~Baltsavias, and K.~Schindler, ``Hyperspectral super-resolution
  with spectral unmixing constraints,'' \emph{Remote Sensing}, vol.~9, no.~11,
  p. 1196, 2017.

\bibitem{liu2019hsr_recovery}
H.~Liu, R.~Wu, and W.-K. Ma, ``Is there any recovery guarantee with coupled
  structured matrix factorization for hyperspectral super-resolution?''
  \emph{{\rm in} Proc. 2019 IEEE Int. Workshop Computational Advances in
  Multi-Sensor Adaptive Processing (CAMSAP)}, 2019.

\bibitem{hsr_recovery_ssp2018}
Q.~Li, W.-K. Ma, and Q.~Wu, ``Hyperspectral super-resolution: Exact recovery in
  polynomial time,'' in \emph{Proc. IEEE Workshop Stat. Signal Process.}, 2018,
  pp. 378--382.

\bibitem{akhtar2014sparse}
N.~Akhtar, F.~Shafait, and A.~Mian, ``Sparse spatio-spectral representation for
  hyperspectral image super-resolution,'' in \emph{European Conference on
  Computer Vision}.\hskip 1em plus 0.5em minus 0.4em\relax Springer, 2014, pp.
  63--78.

\bibitem{wei2015hyperspectral}
Q.~Wei, J.~Bioucas-Dias, N.~Dobigeon, and J.-Y. Tourneret, ``Hyperspectral and
  multispectral image fusion based on a sparse representation,'' \emph{IEEE
  Trans. Geosci. Remote Sens.}, vol.~53, no.~7, pp. 3658--3668, 2015.

\bibitem{kanatsoulis2018hyperspectral}
C.~I. Kanatsoulis, X.~Fu, N.~D. Sidiropoulos, and W.-K. Ma, ``Hyperspectral
  super-resolution: {A} coupled tensor factorization approach,'' \emph{IEEE
  Trans. Signal Process.}, vol.~66, no.~24, pp. 6503--6517, 2018.

\bibitem{li2018fusing}
S.~Li, R.~Dian, L.~Fang, and J.~M. Bioucas-Dias, ``Fusing hyperspectral and
  multispectral images via coupled sparse tensor factorization,'' \emph{IEEE
  Trans. Image Process.}, vol.~27, no.~8, pp. 4118--4130, 2018.

\bibitem{prevost2019coupled}
C.~Pr{\'e}vost, K.~Usevich, P.~Comon, and D.~Brie, ``Coupled tensor low-rank
  multilinear approximation for hyperspectral super-resolution,'' in
  \emph{Proc. IEEE Int. Conf. Acoustics, Speech, Signal Process. (ICASSP)},
  2019.

\bibitem{wright2015coordinate}
S.~J. Wright, ``Coordinate descent algorithms,'' \emph{Math. Program.}, vol.
  151, no.~1, pp. 3--34, 2015.

\bibitem{razaviyayn2013unified}
M.~Razaviyayn, M.~Hong, and Z.-Q. Luo, ``A unified convergence analysis of
  block successive minimization methods for nonsmooth optimization,''
  \emph{SIAM J. Optim.}, vol.~23, no.~2, pp. 1126--1153, 2013.

\bibitem{hong2017iteration}
M.~Hong, X.~Wang, M.~Razaviyayn, and Z.-Q. Luo, ``Iteration complexity analysis
  of block coordinate descent methods,'' \emph{Math. Program.}, vol. 163, no.
  1-2, pp. 85--114, 2017.

\bibitem{xu2013block}
Y.~Xu and W.~Yin, ``A block coordinate descent method for regularized
  multiconvex optimization with applications to nonnegative tensor
  factorization and completion,'' \emph{SIAM J. Imag. Sci.}, vol.~6, no.~3, pp.
  1758--1789, 2013.

\bibitem{xu2017globally}
------, ``A globally convergent algorithm for nonconvex optimization based on
  block coordinate update,'' \emph{J. Sci. Comput.}, vol.~72, no.~2, pp.
  700--734, 2017.

\bibitem{beck2015cyclic}
A.~Beck, E.~Pauwels, and S.~Sabach, ``The cyclic block conditional gradient
  method for convex optimization problems,'' \emph{SIAM J. Optim.}, vol.~25,
  no.~4, pp. 2024--2049, 2015.

\bibitem{beck2018primal}
------, ``Primal and dual predicted decrease approximation methods,''
  \emph{Math. Program.}, vol. 167, no.~1, pp. 37--73, 2018.

\bibitem{beck2017first}
A.~Beck, \emph{First-Order Methods in Optimization}.\hskip 1em plus 0.5em minus
  0.4em\relax Philadelphia, PA, USA: SIAM, 2017, vol.~25.

\bibitem{lange2016mm}
K.~Lange, \emph{MM Optimization Algorithms}.\hskip 1em plus 0.5em minus
  0.4em\relax Philadelphia, PA, USA: SIAM, 2016, vol. 147.

\bibitem{drusvyatskiy2018proximal}
D.~Drusvyatskiy, ``The proximal point method revisited,'' \emph{SIAG/OPT Views
  and News}, vol.~26, 2018.

\bibitem{pmlr-v28-jaggi13}
M.~Jaggi, ``Revisiting {Frank-Wolfe}: Projection-free sparse convex
  optimization,'' in \emph{Proc. Int. Conf. Machine Learning}, 2013, pp.
  427--435.

\bibitem{duchi2008efficient}
J.~Duchi, S.~Shalev-Shwartz, Y.~Singer, and T.~Chandra, ``Efficient projections
  onto the l1-ball for learning in high dimensions,'' in \emph{Proc. Int. Conf.
  Machine Learning}, 2008, pp. 272--279.

\bibitem{agarwal2012fast}
A.~Agarwal, S.~N. Negahban, and M.~J. Wainwright, ``Fast global convergence of
  gradient methods for high-dimensional statistical recovery,'' \emph{The
  Annals of Statistics}, vol.~40, no.~5, pp. 2452--2482, 2012.

\bibitem{Jose12}
J.~Bioucas-Dias, A.~Plaza, N.~Dobigeon, M.~Parente, Q.~Du, P.~Gader, and
  J.~Chanussot, ``Hyperspectral unmixing overview: {G}eometrical, statistical,
  and sparse regression-based approaches,'' \emph{IEEE J. Sel. Topics Appl.
  Earth Observ.}, vol.~5, no.~2, pp. 354--379, 2012.

\bibitem{grippo2000convergence}
L.~Grippo and M.~Sciandrone, ``On the convergence of the block nonlinear
  {Gauss--Seidel} method under convex constraints,'' \emph{Oper. Res. Lett.,},
  vol.~26, no.~3, pp. 127--136, 2000.

\bibitem{fu2016robust}
X.~Fu, K.~Huang, B.~Yang, W.-K. Ma, and N.~D. Sidiropoulos, ``Robust volume
  minimization-based matrix factorization for remote sensing and document
  clustering,'' \emph{IEEE Trans. Signal Process.}, vol.~64, no.~23, pp.
  6254--6268, 2016.

\bibitem{shao2019framework}
M.~Shao, Q.~Li, W.-K. Ma, and A.~M.-C. So, ``A framework for one-bit and
  constant-envelope precoding over multiuser massive {MISO} channels,''
  \emph{{\rm to appear in} IEEE Trans. Signal Process.}, 2019.

\bibitem{lin2015identifiability}
C.-H. Lin, W.-K. Ma, W.-C. Li, C.-Y. Chi, and A.~Ambikapathi, ``Identifiability
  of the simplex volume minimization criterion for blind hyperspectral
  unmixing: The no-pure-pixel case,'' \emph{IEEE Trans. Geosci. Remote Sens.},
  vol.~53, no.~10, pp. 5530--5546, 2015.

\bibitem{calafiore2014optimization}
G.~C. Calafiore and L.~El~Ghaoui, \emph{{O}ptimization {M}odels}.\hskip 1em
  plus 0.5em minus 0.4em\relax Cambridge University Press, 2014.

\bibitem{bertsekas2003convex}
D.~P. Bertsekas, A.~Nedi, and A.~E. Ozdaglar, \emph{{Convex Analysis and
  Optimization}}.\hskip 1em plus 0.5em minus 0.4em\relax Athena Scientific,
  2003.

\bibitem{WALDS_PROTOCOL}
L.~Wald, T.~Ranchin, and M.~Mangolini, ``Fusion of satellite images of
  different spatial resolutions: Assessing the quality of resulting images,''
  \emph{Hotogrammetric Eng. Remote Sens.}, vol.~63, no.~6, pp. 691--699, 1997.

\bibitem{NYOKOYA2016}
\BIBentryALTinterwordspacing
N.~Yokoya and A.~Iwasaki, ``Airborne hyperspectral data over {C}hikusei,''
  Space Application Laboratory, University of Tokyo, Japan, Tech. Rep.
  SAL-2016-05-27, May 2016. [Online]. Available:
  \url{http://park.itc.u-tokyo.ac.jp/sal/hyperdata/TechRepSAL20160527.pdf}
\BIBentrySTDinterwordspacing

\bibitem{dial2003ikonos}
G.~Dial, H.~Bowen, F.~Gerlach, J.~Grodecki, and R.~Oleszczuk, ``{IKONOS}
  satellite, imagery, and products,'' \emph{Remote Sens. Environ.}, vol.~88,
  no. 1-2, pp. 23--36, 2003.

\bibitem{kokaly2017usgs}
R.~F. Kokaly \emph{et~al.}, ``{USGS} spectral library version 7,'' US
  Geological Survey, Tech. Rep., 2017.

\bibitem{vane1993airborne}
G.~Vane, R.~O. Green, T.~G. Chrien, H.~T. Enmark, E.~G. Hansen, and W.~M.
  Porter, ``The airborne visible/infrared imaging spectrometer ({AVIRIS}),''
  \emph{Remote Sens. Environ.}, vol.~44, no. 2-3, pp. 127--143, 1993.

\bibitem{chander2009summary}
G.~Chander, B.~L. Markham, and D.~L. Helder, ``Summary of current radiometric
  calibration coefficients for {Landsat MSS}, {TM}, {ETM+}, and {EO-1 ALI}
  sensors,'' \emph{Remote Sens. Environ.}, vol. 113, no.~5, pp. 893--903, 2009.

\end{thebibliography}

\end{document}